\documentclass[final]{article}

\usepackage[margin=13truemm]{geometry}
\usepackage{authblk}
\usepackage{graphicx}%
\usepackage{multirow}%
\usepackage{amsmath,amssymb,amsfonts}%
\usepackage{amsthm}%
\usepackage{mathrsfs}%
\usepackage[title]{appendix}%
\usepackage{xcolor}%
\usepackage{textcomp}%
\usepackage{manyfoot}%
\usepackage{booktabs}%
\usepackage{algorithm}%
\usepackage{algorithmicx}%
\usepackage{algpseudocode}%
\usepackage{listings}%

\usepackage{mypackage} 
\usepackage{mycommand} 

\theoremstyle{thmstyleone}%
\newtheorem{theorem}{Theorem}
\newtheorem{proposition}{Proposition}

\theoremstyle{thmstyletwo}%
\newtheorem{example}{Example}%
\newtheorem{remark}{Remark}%

\theoremstyle{thmstylethree}%
\newtheorem{definition}{Definition}%
\newtheorem{lemma}{Lemma}

\newtheorem*{conj*}{Conjecture}

\newcommand{\lpnorm}[2]{\left\|#2\right\|_{#1}}
\newcommand{\trdist}[1]{\left\|#1\right\|_{\text{tr}}}
\newcommand{\hh}{\mathcal{H}}
\newcommand{\uu}{\mathcal{U}}

\DeclareMathOperator{\lopp}{\mathbf{L}}
\newcommand{\lop}[1]{\lopp(#1)}

\DeclareMathOperator{\dopp}{\mathbf{D}}
\newcommand{\dop}[1]{\dopp(#1)}

\DeclareMathOperator{\Poss}{\mathbf{Pos}}
\newcommand{\pos}[1]{\Poss(#1)}

\DeclareMathOperator{\unitaryy}{\mathbf{U}}
\newcommand{\unitary}[1]{\unitaryy(#1)}

\newcommand{\idop}{I}
\renewcommand{\tr}[1]{{\rm tr}\left[#1\right]}

\newcommand{\cdim}[1]{\mathbb{C}^{#1}}

\newcommand{\Gcount}[2]{#1\log_{#2}\left(\frac{1}{\varepsilon}\right)}

\newcommand{\prob}{\mathrm{prob}}

\DeclareMathOperator{\CO}{CO}
\DeclareMathOperator{\C}{C}
\DeclareMathOperator{\AQ}{AQ}

\title{Asymptotic Gate Count Bounds for Ancilla-Free Single-Qubit Synthesis with Arithmetic Gates}

\author[1]{Kaoru Sano$^*$}
\author[3]{Hayata Morisaki}
\author[1,2]{Seiseki Akibue}
\affil[1]{\footnotesize NTT Institute for Fundamental Mathematics, Communication Science Laboratories, NTT, Inc.}
\affil[2]{\footnotesize NTT Research Center for Theoretical Quantum Information, NTT, Inc.}
\affil[3]{\footnotesize Graduate School of Engineering Science, The University of Osaka}

\date{}

\begin{document}

\maketitle

\abstract{
We study ancilla-free approximation of single-qubit unitaries $U\in {\rm SU}(2)$ by gate sequences over Clifford+$G$, where $G\in\{T,V\}$ or their generalization. Let $p$ denote the characteristic factor of the gate set (e.g., $p=2$ for $G=T$ and $p=5$ for $G=V$). We prove three asymptotic bounds on the minimum $G$-count required to achieve approximation error at most $\varepsilon$.
First, for Haar-almost every $U$, we show that
 $3\log_{p}(1/\varepsilon)$ $G$-count is both necessary and sufficient; moreover, probabilistic synthesis improves the leading constant to $3/2$.
Second, for unitaries whose ratio of matrix elements lies in a specified number field, $4\log_p(1/\varepsilon)$ $G$-count is necessary. Again, the leading constant can be improved to $2$ by probabilistic synthesis.
Third, there exist unitaries for which the $G$-count per $\log_{p}(1/\varepsilon)$ fails to converge as $\varepsilon\to 0^+$.
These results partially resolve a generalized form of the Ross--Selinger conjecture.

}

\section{Introduction}\label{sec1}

In the era of fault-tolerant quantum computing (FTQC), quantum circuits must be constructed from sequences of elementary gates that are protected from noise owing to quantum error correction (QEC)~\cite{shor1995scheme,steane1996error,calderbank1996good}. The choice of QEC code determines the set of elementary gates: for example, the surface code supports Clifford gates~\cite{kitaev2003fault,fowler2012surface,dennis2002topological}, while the Reed-Muller code allows for multi-controlled-$Z$ gates~\cite{reichardt2005quantum,barg2024geometricstructuretransversallogic}. However, both elementary gate sets are finite for each number of qubits. Importantly, the finiteness of these elementary gate sets is not a byproduct of the specific error correction code employed but rather stems from fundamental constraints imposed by quantum mechanics itself~\cite{eastin2009restrictions}.
To realize universal computation, we often add a few gates in an elementary gate set in compensation for the cost of a procedure for protecting those extra gates from noise, such as magic state distillation~\cite{bravyi2005universal,reichardt2005quantum}, code switching~\cite{anderson2014fault,pogorelov2025codeSwitching}.

This limitation necessitates approximating unitary gates that appear in a circuit, which typically contain continuous parameters, using only sequences consisting of a finite elementary gate set, which is called \textit{approximate unitary synthesis}.
In this paper, we focus on Clifford+$G$ as elementary gate sets, where $G$ can be $T$, $V$, or their generalization, which are the most studied in the context of unitary synthesis.
This setting raises a central question: how can one determine a gate sequence with the minimum number of non-Clifford gates—referred to as the $G$-count—that approximates a target unitary within a specified precision? Although brute-force search can, in principle, identify such optimal sequences, its computational cost grows exponentially with sequence length, making it impractical even for single-qubit unitaries with modest error thresholds such as $\varepsilon \sim 10^{-3}$. To address this, a variety of synthesis algorithms, including suboptimal ones, have been proposed~\cite{dawson2006solovay,BGS13,giles2013exact,kliuchnikov2013fast,ross2014optimal,Gheorghiu2022,Kliuchnikov2023shorterquantum}.

A successful approach to developing a synthesis algorithm has been established following the elucidation of a profound connection between unitary synthesis and number theory~\cite{BGS13,giles2013exact,kliuchnikov2013fast,ross2014optimal,Amy2020numbertheoretic,Kliuchnikov2023shorterquantum,MSA25}. In certain elementary gate sets such as Clifford+$G$ and some gates associated with certain quaternion algebras~\cite{BGS13}, unitaries associated to elementary gate sequences correspond to matrices over specific number fields. The $G$-count relates closely to the height of elements with respect to these fields.

Beyond the development of an algorithm, understanding the asymptotic scaling of the $G$-count associated with synthesizing a fixed target unitary as the acceptable error $\varepsilon$ decreases is crucial for estimating the scaling of spacetime resources required to execute a quantum algorithm on an actual quantum computer.
Previous research has revealed that the number of elementary gates scales $\Theta(\log\left(\frac{1}{\varepsilon}\right))$ for many elementary gate sets, including Clifford+$G$ \cite{HRC02,BG08,BG12} by exploiting their number-theoretic characterization.
For the case of single-qubit unitary synthesis, empirical studies suggest that for most target unitaries, the $G$-count closely follows a lower bound derived from the volume consideration. Additionally, rare \textit{edge cases}~\cite{BGS13,ross2014optimal} exist—also known as \textit{big holes}~\cite{PARZANCHEVSKI2018869}—where the approximation requires substantially larger $G$-count. Ross and Selinger summarize these observations as the following conjecture.

\begin{conj*}[{\cite[Conjecture 8.10]{ross2014optimal}}]
The asymptotic scaling of the $T$-count required to approximate $R_z(\theta):=\exp(-i\theta Z/2)$ within an approximation error $\varepsilon$ is given by
\begin{itemize}
\item $4\log_2\left(\tfrac{1}{\varepsilon}\right)$ if $\tan\frac{\theta}{2} \in \mathbb{Q}(\sqrt{2})$ and $R_z(\theta)$ is not exactly synthesizable,
\item $3\log_2\left(\tfrac{1}{\varepsilon}\right)$ if $\tan\frac{\theta}{2}  \notin \mathbb{Q}(\sqrt{2})$ and $R_z(\theta)$ is not exactly synthesizable.
\end{itemize} 
\end{conj*}

In the case of $V$-count, there is no explicit conjecture, however, a similar behavior has been observed that $V$-count scales as $3\log_5\left(\frac{1}{\varepsilon}\right)$ for most of target unitaries and $4\log_5\left(\frac{1}{\varepsilon}\right)$ for rare cases~\cite{BGS13}.

\begin{figure}[h]
\centering
\includegraphics[width=0.9\textwidth]{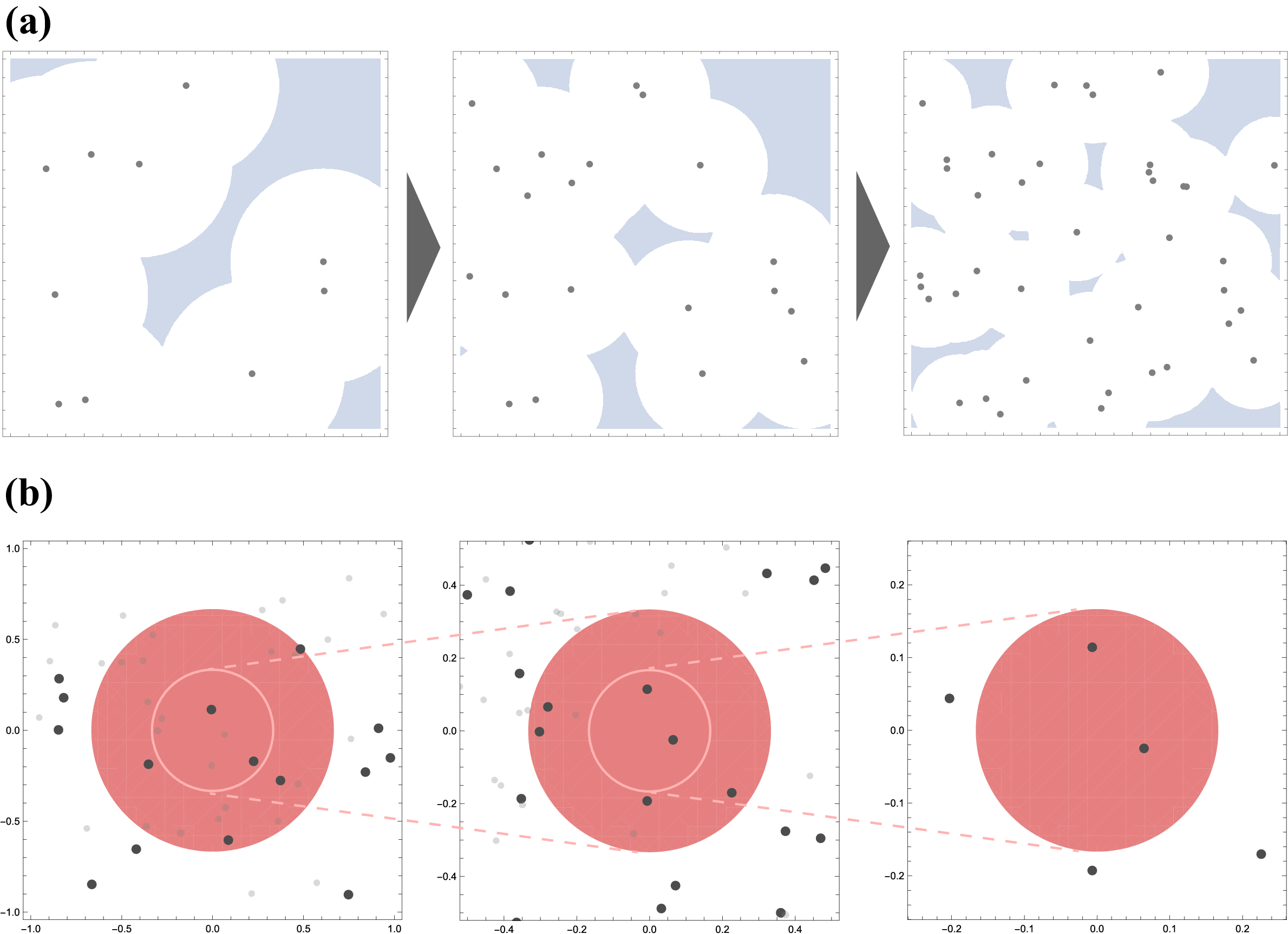}
\caption{Illustration of the difference between (a) the previous research and (b) our research. In both cases, we simultaneously increase the number of exactly synthesizable unitaries (dark gray dots) in compensation for the larger $G$-count and decrease the acceptable error, as shown in the figure from left to right.
(a) \textit{Previous research}~\cite{PARZANCHEVSKI2018869}: The blue region represents target unitaries that exactly synthesizable unitaries cannot approximate within the acceptable error. If the ratio between the approximation error and the number of synthesizable unitaries is appropriately chosen, the area of this region converges to zero. However, this research cannot capture whether a fixed target unitary is contained in the blue region or not.
(b) \textit{Our research}: The focus is on the number of synthesizable unitaries near a specific target unitary (located at the origin), which changes as $6\rightarrow6\rightarrow2$ in the figure. This is illustrated by zooming into the region around the origin. The red disc represents the region around the target unitary for each level of acceptable error. The light gray dots represent the exactly synthesizable unitaries obtained by increasing $G$-count by one.}\label{fig:diagram}
\end{figure}

Despite the considerable body of research conducted by quantum information scientists, theoretical computer scientists, and pure mathematicians, this conjecture remains open.
From a slightly different perspective, Parzanchevski and Sarnak investigated the set of target unitaries that can be approximated within an acceptable error $\varepsilon$ by using a gate sequence with $G$-count of $\C$ when one simultaneously decreases $\varepsilon$ and increases $\C$~\cite{PARZANCHEVSKI2018869}. They proved that the volume of the approximable unitaries approaches unity if $\C\sim 3\log_p\left(\frac{1}{\varepsilon}\right)$; however, the set of the approximable unitaries cannot covers all the single-qubit unitaries unless $\C\geq 4\log_p\left(\frac{1}{\varepsilon}\right)$, where $p=2$ for $G=T$ and $p=5$ for $G=V$.
However, these results do not resolve the conjecture, as they merely demonstrate the existence of a target unitary that is hard to approximate, without specifying what it is. Moreover, even if the volume of the approximable unitaries approaches unity, it is even possible that a particular fixed target unitary is contained in the region of approximable unitaries at specific values of $\varepsilon$ but not contained there at different error levels (see Fig.~\ref{fig:diagram}), which raises the question of whether all target unitaries can be classified simply into two categories as stated in the conjecture.

\section{Results}\label{sec2}

We give some notation on the necessary and sufficient order of $G$-count to approximate a target unitary channel $\mathcal{U}$. We write the precise definition of them in Section~\ref{subsec: Notion for asymptotic G-count}.
We characterize the asymptotic scaling of $G$-count to approximate $\mathcal{U}$ within an approximation error $\varepsilon$ by its upper bound $\Gcount{\overline{\CO}(\mathcal{U},G)}{p}$ and lower bound $\Gcount{\underline{\CO}(\mathcal{U},G)}{p}$ in the limit of $\varepsilon\to0^+$. If they coincide, the \textit{exact $G$-count order} $\CO(\mathcal{U},G)=\overline{\CO}(\mathcal{U},G)=\underline{\CO}(\mathcal{U},G)$ is defined.

We also characterize the asymptotic scaling of $G$-count by its upper bound $\Gcount{\overline{\CO}^{\prob}(\mathcal{U},G)}{p}$, lower bound $\Gcount{\underline{\CO}^{\prob}(\mathcal{U},G)}{p}$, and exact $G$-count order $\CO^{\prob}(\mathcal{U},G)$ if we use probabilistic synthesis—a recent technique that approximates a unitary channel by a mixed unitary channel. Many studies~\cite{PhysRevA.95.042306,hastings2016turning,akibue2024probabilistic} have shown that the probabilistic synthesis typically reduces the approximation error quadratically.

We summarize the main results in Table \ref{table:results}. 
We prove these values, except for the conjectured ones, without any numerical or number-theoretical assumptions.
A $\Gcount{3}{p}$ scaling (or $\Gcount{\tfrac{3}{2}}{p}$ in the probabilistic case) is obtained by combining the theory of optimal probabilistic synthesis~\cite{akibue2024probabilistic} with the covering property of synthesizable unitaries~\cite{PARZANCHEVSKI2018869}. On the other hand, $\Gcount{4}{p}$ scaling (or $\Gcount{2}{p}$ in the probabilistic case) lower bounds emerge from a tight connection between unitary synthesis and Diophantine approximation; leveraging the celebrated Subspace Theorem~\cite{Schm80,Schl77}, we establish the hardness of approximating an edge case $\mathcal{U}$ in a unified framework for both Clifford+$T$ and generalized $V$ gates, and more general gates defined by the arithmetic way.

\begin{table}[ht]
\caption{
Summary of results. Here, a.e. denotes “almost everywhere” with respect to the Haar measure. A unitary $\mathcal{U}$ is said to have $\mathbb{Z}[\sqrt{2}]$-ratio if the ratio of its matrix elements lies in $\mathbb{Z}[\sqrt{2}]$. 
Except for the $\Gcount{6}{p}$-type upper bound, which was previously established by Parzanchevski et al.~\cite{PARZANCHEVSKI2018869}, all the reported values are new. The values marked with an asterisk are conjectural.
}
\label{table:results}
\begin{tabular*}{\textwidth}{@{\extracolsep\fill}lccc}
\toprule%
& \multicolumn{3}{@{}c@{}}{Clifford+$T$}  \\\cmidrule{2-4}%
$G$-count & $\mathcal{U}$ a.e. & $\mathcal{U}$ with $\mathbb{Z}[\sqrt{2}]$-ratio  & Liouville-type $\mathcal{U}$\\
\midrule
$\overline{\CO}(\mathcal{U},T)$ & 3 & $\in[4,6]$ & $\in[4,6]$ \\
$\CO(\mathcal{U},T)$  & 3 & $4^*$ & undefined \\
$\underline{\CO}(\mathcal{U},T)$  & 3 & $\in[4,6]$ & 0 \\
\midrule
$\overline{\CO}^{\prob}(\mathcal{U},T)$  & $3/2$ & $\in[2,3]$ & $\in[2,3]$ \\
$\CO^{\prob}(\mathcal{U},T)$  & $3/2$ & $2^*$ & undefined\\
$\underline{\CO}^{\prob}(\mathcal{U},T)$  & $3/2$ & $\in[2,3]$ & 0 \\
\toprule
\end{tabular*}

\begin{tabular*}{\textwidth}{@{\extracolsep\fill}lccc}
\toprule%
& \multicolumn{3}{@{}c@{}}{Clifford+$V_p$}  \\\cmidrule{2-4}%
$G$-count & $\mathcal{U}$ a.e. & $\mathcal{U}$ with $\mathbb{Z}$-ratio  & Liouville-type $\mathcal{U}$\\
\midrule
$\overline{\CO}(\mathcal{U},V_p)$ & 3 & $\in[4,6]$ & $\in[4,6]$ \\
$\CO(\mathcal{U},V_p)$  & 3 & $4^*$ & undefined \\
$\underline{\CO}(\mathcal{U},V_p)$  & 3 & $\in[4,6]$ & 0 \\
\midrule
$\overline{\CO}^{\prob}(\mathcal{U},V_p)$  & $3/2$ & $\in[2,3]$ & $\in[2,3]$ \\
$\CO^{\prob}(\mathcal{U},V_p)$  & $3/2$ & $2^*$ & undefined\\
$\underline{\CO}^{\prob}(\mathcal{U},V_p)$  & $3/2$ & $\in[2,3]$ & 0 \\
\toprule
\end{tabular*}

\end{table}

\section{Preliminaries}
In this section, we summarize basic notations used throughout the paper. $U(d)$ is the set of $d$ by $d$ unitary matrices, and $\SU(d):=\{U\in U(d):\det U=1\}$.
 Note that we consider only finite-dimensional Hilbert spaces. In particular, a two-dimensional Hilbert space $\cdim{2}$ is called a qubit.
The $\lop{\hh}$ and $\pos{\hh}$ represent the set of linear operators and positive semidefinite operators on Hilbert space $\hh$, respectively. $\unitary{\hh}$ represents the set of unitary operators.
$\idop\in\pos{\hh}$ represents the identity operator.
The $\dop{\hh}:=\left\{\rho\in\pos{\hh}:\tr{\rho}=1\right\}$ represents the set of quantum states. Any physical transformation of the quantum state can be represented by a completely positive and trace-preserving (CPTP) linear mapping $\Gamma:\lop{\hh_1}\longrightarrow\lop{\hh_2}$.
 
The trace distance $\trdist{\rho-\sigma}$ of two quantum states $\rho,\sigma\in\dop{\hh}$ is defined as $\trdist{M}:=\frac{1}{2}\tr{\sqrt{MM^\dag}}$ for $M\in\lop{\hh}$. It represents the maximum total variation distance between probability distributions obtained from measurements performed on two quantum states.

The distance measuring the distinguishability of two CPTP mappings
\[
    \mathcal{A},\mathcal{B}:\lop{\hh}\longrightarrow\lop{\hh}
\]
corresponding to the trace distance is the diamond distance $d(\mathcal{U},\mathcal{V})$ defined by 
\[
    d(\mathcal{U},\mathcal{V}) \coloneqq \max_{\rho\in\dop{\hh\otimes\hh}}\trdist{((\mathcal{A}-\mathcal{B})\otimes \id)(\rho)},
\]
where $\id$ represents the identity mapping acting on $\hh$. Note that the diamond distance can be regarded as a norm over the vector space spanned by CPTP mappings.

\subsection{Deterministic and probabilistic unitary synthesis}
For a unitary operator $U\in\unitary{\hh}$, we associated the CPTP map $\uu:\lop{\hh}\longrightarrow\lop{\hh}$ defined by
\begin{equation}
    \uu(\rho)=U\rho U^\dag,
\end{equation}
which describes the physical time evolution of a quantum state $\rho$ under the unitary transformation $U$.
A CPTP map expressed in this form is referred to as a unitary channel.
We sometimes denote $\mathcal{U}_U$ with a subscript to emphasize the underlying unitary operator $U$ that generates the transformation $\mathcal{U}_U$.
Note that $d(\mathcal{U},\mathcal{V}_1\circ\mathcal{V}_2)=d(\mathcal{V}_1^{-1}\circ\mathcal{U},\mathcal{V}_2)=d(\mathcal{U}\circ\mathcal{V}_2^{-1},\mathcal{V}_1)$ holds for any unitary channels $\mathcal{U}$, $\mathcal{V}_1$ and $\mathcal{V}_2$.

A more general CPTP map $\mathcal{E}$, realizable by probabilistical sampling of unitary channels $\{\mathcal{U}_x\}_x$, is called a \textit{mixed unitary channel} and is represented by
\begin{equation}
 \mathcal{E}(\rho)=\sum_{x}p(x)\mathcal{U}_x(\rho)=\sum_{x}p(x)U_x\rho U_x^\dag.
\end{equation}

For a metric space $(X,d)$ and a subset $S\subseteq X$, $S$ is called an \textit{$\varepsilon$-covering} of $X$ if $\sup_{t\in X}\inf_{s\in S}d(s,t)\leq\varepsilon$.
In this work, we basically consider $X$ to be either the set of unitary channels or a $\delta$-ball centered at a unitary channel $\mathcal{U}$, defined as $\{\mathcal{V}:d(\mathcal{U},\mathcal{V})\leq\delta\}$, where the diamond distance gives the metric.

In this work, we focus on single-qubit unitary operators, which can be represented as unitary matrices in $U(2)$ with respect to a computational basis. Fixing the computational basis, we henceforth identify each unitary operator with its matrix representation.
A unitary operator is often referred to as a gate in the context of unitary synthesis.

In deterministic unitary synthesis, the goal is to find a single unitary channel $\mathcal{V}$ that can be exactly realized by using an elementary gate sequence and serves as an approximation to a target unitary channel $\mathcal{U}$. To quantify the approximation error, we employ the diamond distance $d(\mathcal{U},\mathcal{V})$, which captures the fundamental distinguishability between CPTP maps.
Although the diamond norm between two unitary channels generally lacks a simple analytical expression, for the case of single-qubit unitaries, it admits a closed form due to Akibue et al.~\cite{akibue2024probabilistic} (see also \cite[Proposition 2.1]{MSA25}):
\begin{equation}
\label{eq:diamond_distance_1Q}
d(\mathcal{U},\mathcal{V})=\sqrt{1-\left(\frac{1}{2}\left|\tr{U^\dag V}\right|\right)^2}. 
\end{equation}

When unitary channels $\{\mathcal{V}_x\}_x$ can be exactly implemented by using elementary gate sequences, a mixed unitary channel $\sum_xp(x)\mathcal{V}_x$ can be realized by probabilistically sampling the label $x$ according to the probability distribution $p(x)$ and executing the corresponding gate sequence. The only additional cost comes from sampling and adaptively switching the gate sequence, with no post-processing required. This motivates us to consider probabilistic unitary synthesis, which seeks a mixed unitary channel to approximate $\mathcal{U}$.

More precisely, the goal of probabilistic synthesis is to find set of unitary channels $\{\mathcal{V}_x\}_x$, each exactly realized by using an elementary gate sequence, together with a probability distribution $p(x)$ such that $\sum_xp(x)\mathcal{V}_x$ serves as an approximation to a target unitary channel $\mathcal{U}$. The approximation error is again quantified using the diamond distance.
Counterintuitively, probabilistic synthesis can substantially reduce the approximation error, even though a unitary channel is not itself a probabilistic mixture of distinct unitaries.
Akibue et al.~\cite{akibue2024probabilistic} have derived the following two statements to characterize the optimal probabilistic synthesis.
\begin{lemma}\cite[Theorem 4.3]{akibue2024probabilistic}
\label{lemma:probsyntherrorbound}
 For a target single-qubit unitary channel $\mathcal{U}$ and a finite set $\{\mathcal{V}_x\}_x$ of single-qubit unitary channels, it holds that
\begin{equation}
\left( \min_xd(\mathcal{U},\mathcal{V}_x)\right)^2\leq\min_p d\left(\mathcal{U},\sum_xp(x)\mathcal{V}_x\right)\leq \left(\max_{\mathcal{U}} \min_xd(\mathcal{U},\mathcal{V}_x)\right)^2.
\end{equation}
\end{lemma}

\begin{lemma}\cite[Lemma 5.3]{akibue2024probabilistic}
\label{lemma:probsynthsupport}
 For a non-negative real number $\varepsilon\geq0$ and a target single-qubit unitary channel $\mathcal{U}$, if $\{\mathcal{V}_x\}_x$ is a finite $\varepsilon$-covering of the set of single-qubit unitary channels, i.e., $\max_{\mathcal{U}} \min_xd(\mathcal{U},\mathcal{V}_x)\leq\varepsilon$, then
\begin{equation}
\min_{\hat{p}} d\left(\mathcal{U},\sum_x\hat{p}(x)\mathcal{V}_x\right)=\min_p d\left(\mathcal{U},\sum_xp(x)\mathcal{V}_x\right)
\end{equation}
holds, where $\hat{p}$ has its support on $\hat{X}:=\{x:d(\mathcal{U},\mathcal{V}_x)\leq2\varepsilon\}$.
\end{lemma}

By combining these two lemmas, we obtain the following proposition, which plays a central role in the analysis of $G$-count in probabilistic synthesis.
\begin{proposition}
\label{prop:probsynth}
 For a non-negative real number $\varepsilon\geq0$ and a target single-qubit unitary channel $\mathcal{U}$, if $\{\mathcal{V}_x\}_x$ is a finite $\varepsilon$-covering of the $(2\varepsilon)$-ball centered at $\mathcal{U}$, then it holds that
\begin{equation}
\label{eq:probsynth}
\left( \min_xd(\mathcal{U},\mathcal{V}_x)\right)^2\leq\min_p d\left(\mathcal{U},\sum_xp(x)\mathcal{V}_x\right)\leq \varepsilon^2.
\end{equation}
\end{proposition}
\begin{proof}
 Since the first inequality in Eq.~\eqref{eq:probsynth} is a direct consequence of Lemma \ref{lemma:probsyntherrorbound}, we show the second one.
 Let $\{\mathcal{V}'_y\}_y$ be a finite $\varepsilon$-covering of the complement of the $(2\varepsilon)$-ball centered at $\mathcal{U}$ and $d(\mathcal{V}'_y,\mathcal{U})>2\varepsilon$ for any $y$. Then, $\{\mathcal{V}_x\}_x\cup\{\mathcal{V}'_y\}_y$ is an $\varepsilon$-covering of the set of single-qubit unitary channels. By using Lemma \ref{lemma:probsyntherrorbound} and Lemma \ref{lemma:probsynthsupport}, we obtain
 \begin{eqnarray}
 \min_p d\left(\mathcal{U},\sum_xp(x)\mathcal{V}_x\right)= \min_q d\left(\mathcal{U},\sum_xq(x)\mathcal{V}_x+\sum_yq(y)\mathcal{V}'_y\right)\leq\varepsilon^2,
\end{eqnarray}
where $q$ is a probability distribution such that $\sum_xq(x)+\sum_yq(y)=1$.
\end{proof}

\subsection{Strong approximation theory}
Let $S_{\calO_K}(n)$ be the set of integer points $(\alpha,\beta,\gamma,\delta)\in\calO_K^4$ satisfying
\begin{equation}
    \alpha^2+\beta^2+\gamma^2+\delta^2=n,
\end{equation}
where we assume $\calO_K$ is either $\mathbb{Z}$ or $\mathbb{Z}[\sqrt{2}]:=\{a+b\sqrt{2}:a,b\in\mathbb{Z}\}$ in this paper.
We consider a unitary channel $\mathcal{U}(\alpha,\beta,\gamma,\delta)$ associated with an integer point $(\alpha,\beta,\gamma,\delta)$ as follows:
\begin{equation}
    \mathcal{U}(\alpha,\beta,\gamma,\delta)(\rho)=\frac{1}{\alpha^2+\beta^2+\gamma^2+\delta^2}\begin{pmatrix}
        \alpha+i\beta&-\gamma+i\delta\\
        \gamma+i\delta&\alpha-i\beta
    \end{pmatrix}\rho \begin{pmatrix}
        \alpha+i\beta&-\gamma+i\delta\\
        \gamma+i\delta&\alpha-i\beta
    \end{pmatrix}^\dag.
\end{equation}
Parzanchevski et al. have established the following propositions concerning the approximation of points on the three-dimensional sphere by integer points lying on it~\cite{PARZANCHEVSKI2018869}.

\begin{proposition}\cite[Proposition 3.1]{PARZANCHEVSKI2018869}
\label{prop:uncovered_area}
    There exists a positive number $C>0$ such that 
    \begin{itemize}
        \item for a single-qubit unitary channel $\mathcal{V}$ sampled randomly with respect to the Haar measure, the probability that $\mathcal{V}$ cannot be approximated by unitary channels associated with $S_{\mathbb{Z}[\sqrt{2}]}(2^k)$ is at most $C\frac{k^2}{2^{2k}\varepsilon^3}$, i.e.,
        \begin{equation}
            \mu\left(\{\mathcal{V}:\forall(\alpha,\beta,\gamma,\delta)\in S_{\mathbb{Z}[\sqrt{2}]}(2^k),d(\mathcal{V},\mathcal{U}(\alpha,\beta,\gamma,\delta))>\varepsilon\}\right)\leq C\frac{k^2}{2^{2k}\varepsilon^3},
        \end{equation}
        and
        
        \item for a single-qubit unitary channel $\mathcal{V}$ sampled randomly with respect to the Haar measure, the probability that $\mathcal{V}$ cannot be approximated by unitary channels associated with $S_{\mathbb{Z}}(p^k)$ is at most $C\frac{k^2}{p^{k}\varepsilon^3}$, i.e.,
        \begin{equation}
            \mu\left(\{\mathcal{V}:\forall(\alpha,\beta,\gamma,\delta)\in S_{\mathbb{Z}}(p^k),d(\mathcal{V},\mathcal{U}(\alpha,\beta,\gamma,\delta))>\varepsilon\}\right)\leq C\frac{k^2}{p^{k}\varepsilon^3}.
        \end{equation}
    \end{itemize}
\end{proposition}

\begin{proposition}\cite[Corollary 3.2]{PARZANCHEVSKI2018869}
\label{prop:coveringcond}
    There exists a positive number $C>0$ such that 
    \begin{itemize}
        \item $\left\{\mathcal{U}\left(\alpha,\beta,\gamma,\delta\right):(\alpha,\beta,\gamma,\delta)\in S_{\mathbb{Z}[\sqrt{2}]}(2^k)\right\}$ is an $\varepsilon$-covering of the set of unitary channels if $\frac{k}{2^k}\leq C\varepsilon^3$, and

        \item for any odd prime $p$, $\left\{\mathcal{U}\left(\alpha,\beta,\gamma,\delta\right):(\alpha,\beta,\gamma,\delta)\in S_{\mathbb{Z}}(p^k)\right\}$ is an $\varepsilon$-covering of the set of unitary channels if $\frac{k}{p^{\frac{k}{2}}}\leq C\varepsilon^3$.
    \end{itemize}
\end{proposition}

\section{Results}

\subsection{Elementary gate sets}
We focus on the following two classes of elementary gate sets, which are among the most widely used in the field of unitary synthesis.
Recall that the set $\mathcal{C}$ of single-qubit Clifford gates can be generated by $S$ and $H$ gates, defined as
\begin{equation}
     S=\begin{pmatrix}
        1&0\\
        0&i
    \end{pmatrix},\quad 
    H=\frac{1}{\sqrt{2}}\begin{pmatrix}
        1&1\\
        1&-1
    \end{pmatrix}.
\end{equation}
It is known that the size of the set $\{\mathcal{U}_g:g\in\mathcal{C}\}$ of unitary channels corresponding to single-qubit Clifford gates is $24$.
\begin{itemize}
 \item \textbf{Clifford+$T$} is an elementary gate set consisting of  $\mathcal{C}$ and
     \begin{equation}
        T=\begin{pmatrix}
        1&0\\
        0&\zeta_8
    \end{pmatrix},   
     \end{equation}
    where we write $\zeta_n$ for $\exp(2\pi i/n)$. 
    Matsumoto and Amano have shown that any unitary operator generated by Clifford+$T$ can be represented by a canonical form $(T | \varepsilon)(HT | SHT)^* \mathcal{C}$~\cite{matsumoto2008representationquantumcircuitsclifford}.

    It is known that
    \begin{equation}
    \label{eq:Tcount_IntegerPoint}
        \left\{\mathcal{U}\left(\alpha,\beta,\gamma,\delta\right):(\alpha,\beta,\gamma,\delta)\in S_{\mathbb{Z}[\sqrt{2}]}(2^k)\right\}\subseteq
        \{\mathcal{U}:\C(\mathcal{U},T,0)\leq2k+1\},
    \end{equation}
    where $\C(\mathcal{U},T,0)$ is the minimum number of $T$ gates to synthesize $\mathcal{U}$ by using Clifford+$T$~\cite{giles2019remarksmatsumotoamanosnormal}.
    Note that $\C(\mathcal{U},T,0)$ is defined as $\infty$ is $\mathcal{U}$ is not exactly synthesizable.

 \item \textbf{Clifford+$V_p$} is an elementary gate set consisting of	 $\mathcal{C}$ and
\begin{equation}
 \frac{1}{ \sqrt{p}}(\alpha\idop + \beta i Z - \gamma iY + \delta iX),
 \end{equation}
 where $p$ is an odd prime, $X$, $Y$, $Z$ are Pauli matrices, and integers $\alpha,\beta,\gamma,\delta\in\mathbb{Z}$ satisfy $\alpha^2+\beta^2+\gamma^2+\delta^2=p$.
Note that this is a generalization of the $V$ gates, which corresponds to the case $p=5$. Since any Clifford gate commutes with the set of Pauli matrices, any unitary operator generated by Clifford+$V_p$ can be represented by a canonical form $V_p^{(i_1)}V_p^{(i_2)}\cdots V_p^{(i_r)}\mathcal{C}$, where
$\{V_p^{(i)}\}_{i=1}^{p+1}$ is a set of representatives of
$\left\{\frac{1}{ \sqrt{p}}(\alpha\idop + \beta i Z - \gamma iY + \delta iX):\alpha^2+\beta^2+\gamma^2+\delta^2=p\right\}/\mathcal{P}$ and $i_{v+1}$ is chosen so as to satisfy $V_p^{(i_v)}V_p^{(i_{v+1})}\neq\idop$ for $1\leq v \leq r-1$, where $\mathcal{P}=\{    \pm \idop, \pm iX, \pm iY, \pm iZ
\}$. A detailed decomposition into this canonical form is shown by the authors~\cite{SMA25}.

    It is known that
    \begin{equation}
    \label{eq:Vcount_IntegerPoint}
        \left\{\mathcal{U}\left(\alpha,\beta,\gamma,\delta\right):(\alpha,\beta,\gamma,\delta)\in S_{\mathbb{Z}}(p^k)\right\}\subseteq
        \{\mathcal{U}:\C(\mathcal{U},V_p,0)\leq k\},
    \end{equation}
    where $\C(\mathcal{U},V_p,0)$ is the minimum number of $V_p$ gates to synthesize $\mathcal{U}$ by using Clifford+$V_p$ (\cite{BGS13} for the case $p=5$ and \cite{SMA25} for general $p$).
    Note that $\C(\mathcal{U},V_p,0)$ is defined as $\infty$ is $\mathcal{U}$ is not exactly synthesizable.
\end{itemize}

\subsection{Notions for asymptotic $G$-count}
\label{subsec: Notion for asymptotic G-count}
Since the non-Clifford gate $G$ (which in our case is either $T$ or $V_p$) is more challenging to implement than Clifford gates, we introduce notions to analyze the asymptotic behavior of the $G$-count for approximating a target unitary channel $\mathcal{U}$.

In deterministic unitary synthesis, the following quantities, referred to as the necessary $G$-count order and the sufficient one, characterize the asymptotic $G$-count.
\begin{eqnarray}
    \underline{\CO}(\mathcal{U},G) &:=& \sup\left\{t\in\mathbb{R}:\exists\varepsilon_0>0,\forall\varepsilon\in(0,\varepsilon_0),\C(\mathcal{U},G,\varepsilon)\geq t\log_p\left(\frac{1}{\varepsilon}\right)\right\},\\
    \overline{\CO}(\mathcal{U},G) &:=& \inf\left\{t\in\mathbb{R}:\exists\varepsilon_0>0,\forall\varepsilon\in(0,\varepsilon_0), {\C}(\mathcal{U},G,\varepsilon)\leq t\log_p\left(\frac{1}{\varepsilon}\right)\right\},
\end{eqnarray}
where we set $p=2$ in the case $G=T$, and let $\C(\mathcal{U},G,\varepsilon)$ denote the $G$-count of $\mathcal{U}$ in $\varepsilon$-approximation; that is, the minimum number of $G$ gates required to construct a Clifford+$G$ unitary channel $\mathcal{V}$ satisfying $d(\mathcal{U},\mathcal{V})\leq\varepsilon$.
When $\overline{\CO}(\mathcal{U}, G)$ and  $\underline{\CO}(\mathcal{U}, G)$ coincide, we refer to their common value as the \textit{exact $G$-count order} of $\mathcal{U}$ in deterministic synthesis and denote it as $\CO(\mathcal{U}, G)$; otherwise, the exact $G$-count order is said to be undefined.

In probabilistic unitary synthesis, the following quantities characterize the asymptotic $G$-count.
\begin{eqnarray}
    \underline{\CO}^{\prob}(\mathcal{U},G) &:=& \sup\left\{t\in\mathbb{R}:\exists\varepsilon_0>0,\forall\varepsilon\in(0,\varepsilon_0),\C^{\prob}(\mathcal{U},G,\varepsilon)\geq t\log_p\left(\frac{1}{\varepsilon}\right)\right\},\\
    \overline{\CO}^{\prob}(\mathcal{U},G) &:=& \inf\left\{t\in\mathbb{R}:\exists\varepsilon_0>0,\forall\varepsilon\in(0,\varepsilon_0),\C^{\prob}(\mathcal{U},G,\varepsilon)\leq t\log_p\left(\frac{1}{\varepsilon}\right)\right\},
\end{eqnarray}
where let $\C^{\prob}(\mathcal{U},G,\varepsilon)$ be the minimum number $t$ of $G$ gates such that there exist a probability distribution $p(x)$ and a set  $\{\mathcal{V}_x\}_x$ of Clifford+$G$ unitary channels each of whose $G$-count is not greater than $t$ satisfying $d\left(\mathcal{U},\sum_xp(x)\mathcal{V}_x\right)\leq\varepsilon$.
When $\underline{\CO}^{\prob}(\mathcal{U}, G)$ and $\overline{\CO}^{\prob}(\mathcal{U}, G)$ coincide, we again refer to their common value as the exact $G$-count order of $\mathcal{U}$ in probabilistic synthesis and denote it as $\CO^{\prob}(\mathcal{U}, G)$; otherwise, the exact $G$-count order is said to be undefined.

By using Proposition \ref{prop:probsynth}, we obtain the following relationship between deterministic and probabilistic $G$-count.
\begin{proposition}
\label{prop:Gcountprobdet}
 For any $G\in\{T\}\cup\{V_p\}_{p:{\rm odd\ prime}}$ and any single-qubit unitary channel $\mathcal{U}$, it holds that
\begin{eqnarray}
\label{eq:Gcountprobdet1}
 \frac{1}{2}\underline{\CO}(\mathcal{U},G)\leq \underline{\CO}^{\prob}(\mathcal{U},G)\leq \underline{\CO}(\mathcal{U},G),\quad  \frac{1}{2}\overline{\CO}(\mathcal{U},G)\leq \overline{\CO}^{\prob}(\mathcal{U},G).
\end{eqnarray}
\end{proposition}
\begin{proof}
 For any $t<\underline{\CO}(\mathcal{U},G)$, there exist $\varepsilon_0>0$ such that for any $\varepsilon\in(0,\varepsilon_0)$, $\C(\mathcal{U},G,\varepsilon)\geq \Gcount{t}{p}$.
 This implies that $d(\mathcal{U},\mathcal{V})>\varepsilon$ for any Clifford+$G$ unitary channel $\mathcal{V}$ whose $G$-count is less than $\Gcount{t}{p}$.
 By using Proposition \ref{prop:probsynth}, we obtain that $d\left(\mathcal{U},\sum_xp(x)\mathcal{V}_x\right)>\varepsilon^2$ for any probability distribution $p(x)$ and any set $\{\mathcal{V}_x\}_x$ of Clifford+$G$ unitary channels each of whose $G$-count is less than $\Gcount{t}{p}$.
 This implies that $ \underline{\CO}^{\prob}(\mathcal{U},G) \geq\frac{t}{2}$. Thus, we obtain the first inequality of Eq.~\eqref{eq:Gcountprobdet1}.
 The second inequality of Eq.~\eqref{eq:Gcountprobdet1} can be verified by definition.
 
  For any $t>\overline{\CO}^{\prob}(\mathcal{U},G)$, there exist $\varepsilon_0>0$ such that for any $\varepsilon\in(0,\varepsilon_0)$, $\C^{\prob}(\mathcal{U},G,\varepsilon)\leq \Gcount{t}{p}$. This implies that there exist probability distribution $p(x)$ and a set $\{\mathcal{V}_x\}_x$ of Clifford+$G$ unitary channels each of whose $G$-count is not greater than $\Gcount{t}{p}$ such that $d\left(\mathcal{U},\sum_xp(x)\mathcal{V}_x\right)\leq\varepsilon$.
  By using Proposition \ref{prop:probsynth}, we obtain that there exists a Clifford+$G$ unitary channel $\mathcal{V}$ whose $G$-count is not greater than $\Gcount{t}{p}$ such that $d(\mathcal{U},\mathcal{V})\leq\sqrt{\varepsilon}$.
  This implies that $\CO(\mathcal{U},G)\leq 2t$. Thus, we obtain the last inequality of Eq.~\eqref{eq:Gcountprobdet1}.
\end{proof}
This proposition implies that $\underline{\CO}^{\prob}(\mathcal{U},G) \geq 3/2$ for $\mathcal{U}$ a.e., $\underline{\CO}^{\prob}(\mathcal{U},G) \geq 2$ for $\mathcal{U}$ with $\mathbb{Z}[\sqrt{2}]$-ratio or $\mathbb{Z}$-ratio, and $\underline{\CO}^{\prob}(\mathcal{U},G) = 0$ with $\overline{\CO}^{\prob}(\mathcal{U},G) \geq 2$ for Liouville-type $\mathcal{U}$ in Table~\ref{table:results}, provided the deterministic $G$-count results are established.

\begin{proposition}
\label{prop:Gcountprobupper}
 For any $G\in\{T\}\cup\{V_p\}_{p:{\rm odd\ prime}}$ and any single-qubit unitary channel $\mathcal{U}$, it holds that
\begin{eqnarray}
 \overline{\CO}^{\prob}(\mathcal{U},G)\leq3.
\end{eqnarray}
\end{proposition}
\begin{proof}
    Since Proposition \ref{prop:coveringcond} implies that the set $\{\mathcal{V}_x\}_x$ of exactly synthesizable unitary channels whose $G$-count $\C$ satisfies
    $\C\geq\Gcount{6}{p}+2\log_p\C+c$ with some constant $c$ forms an $\varepsilon$-covering of the set of unitary channels. By using Lemma ~\ref{lemma:probsyntherrorbound}, we find that the probabilistic mixture of $\{\mathcal{V}_x\}_x$ can approximate any $\mathcal{U}$ within approximation error $\varepsilon^2$. By definition, this completes the proof.
\end{proof}

\subsection{Theorems on asymptotic $G$-count}

\begin{theorem}\label{thm: G-rate equal 3}
Let $G$ be either $T$ or $V_p$ with an odd prime $p$.
    For a randomly sampled single-qubit unitary channel $\mathcal{U}$ with respect to the Haar measure, $\CO(\mathcal{U},G)=3$ and $\CO^{\prob}(\mathcal{U},G)=\frac{3}{2}$ with probability $1$.    
\end{theorem}
To show this theorem, we first show the following lemmas.
\begin{lemma}
\label{lemma:aelower}
    Let $G$ be either $T$ or $V_p$ with an odd prime $p$.
    For a randomly sampled single-qubit unitary channel $\mathcal{U}$, $\underline{\CO}(\mathcal{U},G)\geq3$ with probability $1$.
\end{lemma}
We use volume considerations differently, as in \cite{Sel15}, to prove this Proposition. This is because even if we can show that $\forall\varepsilon,\exists\mathcal{U},\C(\mathcal{U},T,\varepsilon)\geq3\log_2\left(\frac{1}{\varepsilon}\right)-c$ as \cite{Sel15}, it is not trivial that $\exists\mathcal{U},\forall\varepsilon,\C(\mathcal{U},T,\varepsilon)\geq3\log_2\left(\frac{1}{\varepsilon}\right)-c$. 
\begin{proof}
    If a target unitary channel $\mathcal{U}$ satisfies $\underline{\CO}(\mathcal{U},G)\leq3-2\delta$ with $\delta\in(0,1)$, we can verify 
    \begin{equation}
        \mathcal{U}\in\bigcap_{n\in\mathbb{N}}\bigcup_{\varepsilon\in\left(0,\varepsilon_n\right)}E\left(\varepsilon\right),\ \ E(\varepsilon):=\left\{\mathcal{U}:\C(\mathcal{U},G,\varepsilon)\leq\Gcount{(3-\delta)}{p}\right\},
    \end{equation}
    where $\varepsilon_n$ is defined as $(3-\delta)\log_p\left(\frac{1}{\varepsilon_n}\right)=n\Leftrightarrow\varepsilon_n^{-1}=p^{\frac{n}{3-\delta}}$ for $n\in\mathbb{N}$.

    Since $\mu(E(\varepsilon_t))\leq c\varepsilon_t^3\cdot p^t=c\left(p^{-\frac{\delta}{3-\delta}}\right)^t= cr^t$ with some constant $c>0$ and $r\in(0,1)$ due the canonical forms of Clifford+$G$ sequences, we can obtain
    \begin{equation}
        \mu\left(\bigcap_{n\in\mathbb{N}}\bigcup_{\varepsilon\in\left(0,\varepsilon_n\right)}E\left(\varepsilon\right)\right)
        \leq
        \mu\left(\bigcup_{\varepsilon\in\left(0,\varepsilon_n\right)}E\left(\varepsilon\right)\right)
        \leq
        \mu\left(\bigcup_{t\geq n}E\left(\varepsilon_t\right)\right)
        \leq\frac{c}{1-r}r^n
    \end{equation} 
    for any $n\in\mathbb{N}$, where we used $\bigcup_{\varepsilon\in\left(0,\varepsilon_n\right)}E\left(\varepsilon\right)\subseteq\cup_{t\geq n}E(\varepsilon_t)$ to derive the second inequality. This completes the proof.
    
    \end{proof}

\begin{lemma}
\label{lemma:aeupper}
    Let $G$ be either $T$ or $V_p$ with an odd prime $p$.
    For a randomly sampled single-qubit unitary channel $\mathcal{U}$, $\overline{\CO}(\mathcal{U},G)\leq3$ with probability $1$.
\end{lemma}
\begin{proof}
        If a target unitary channel $\mathcal{U}$ satisfies $\overline{\CO}(\mathcal{U},G)\geq3+2\delta$ with $\delta>0$, we find
    \begin{equation}
        \mathcal{U}\in\bigcap_{n\in\mathbb{N}}\bigcup_{\varepsilon\in\left(0,\varepsilon_n\right)}E\left(\varepsilon\right),\ \ E(\varepsilon):=\left\{\mathcal{U}:\C(\mathcal{U},G,\varepsilon)\geq\Gcount{(3+\delta)}{p}\right\},
    \end{equation}
    where $\varepsilon_n$ is defined as $(3+\delta)\log_p\left(\frac{1}{\varepsilon_n}\right)=n\Leftrightarrow\varepsilon_n^{-1}=p^{\frac{n}{3+\delta}}$ for $n\in\mathbb{N}$.

    Observe that if $\mathcal{V}\in E(\varepsilon_t)$, $d(\mathcal{V},\mathcal{U})>\varepsilon_t$ for any $\mathcal{U}$ whose $G$-count is less than $t(\in\mathbb{N})$.
    When $G=T$, Eq.~\eqref{eq:Tcount_IntegerPoint} implies that $d(\mathcal{V},\mathcal{U}(\alpha,\beta,\gamma,\delta))>\varepsilon_t$ for any integer point $(\alpha,\beta,\gamma,\delta)\in S_{\mathbb{Z}[\sqrt{2}]}(2^{\frac{t}{2}-1})$.
    When $G=V_p$, Eq.~\eqref{eq:Vcount_IntegerPoint} implies that $d(\mathcal{V},\mathcal{U}(\alpha,\beta,\gamma,\delta))>\varepsilon_t$ for any integer point $(\alpha,\beta,\gamma,\delta)\in S_{\mathbb{Z}}(p^{t-1})$.  
    In both cases, Proposition \ref{prop:uncovered_area} implies
    $\mu(E(\varepsilon_t))\leq c\frac{t^2}{p^t\varepsilon_t^3}=ct^2\left(p^{-\frac{\delta}{3+\delta}}\right)^t\leq cr^t$ with some constant $c>0$ and $r\in(0,1)$. 
    Thus, we obtain
    \begin{equation}                        \mu\left(\bigcap_{n\in\mathbb{N}}\bigcup_{\varepsilon\in\left(0,\varepsilon_n\right)}E\left(\varepsilon\right)\right)
        \leq
        \mu\left(\bigcup_{\varepsilon\in\left(0,\varepsilon_n\right)}E\left(\varepsilon\right)\right)
        \leq
        \mu\left(\bigcup_{t\geq n}E\left(\varepsilon_t\right)\right)
        \leq\frac{c}{1-r}r^n
    \end{equation} 
    for any $n\in\mathbb{N}$, where we used $\bigcup_{\varepsilon\in\left(0,\varepsilon_n\right)}E\left(\varepsilon\right)\subseteq\cup_{t\geq n}E(\varepsilon_t)$ to derive the second inequality. This completes the proof.
\end{proof}

\begin{lemma}
\label{lemma:aeupper_prob}
    Let $G$ be either $T$ or $V_p$ with an odd prime $p$.
    For a randomly sampled single-qubit unitary channel $\mathcal{U}$, $\overline{\CO}^{\prob}(\mathcal{U},G)\leq\frac{3}{2}$ with probability $1$.
\end{lemma}
\begin{proof}
    If a target unitary channel $\mathcal{U}$ satisfies $\overline{\CO}^{\prob}(\mathcal{U},G)\geq\frac{3+2\delta}{2}$ with $\delta>0$, we find
    \begin{equation}
        \mathcal{U}\in\bigcap_{n\in\mathbb{N}}\bigcup_{\varepsilon\in\left(0,\varepsilon_n\right)}E\left(\varepsilon\right),\ \ E(\varepsilon):=\left\{\mathcal{U}:\C^{\prob}(\mathcal{U},G,\varepsilon)\geq\Gcount{\frac{3+\delta}{2}}{p}\right\},
    \end{equation}
    where $\varepsilon_n$ is defined as $\frac{3+\delta}{2}\log_p\left(\frac{1}{\varepsilon_n}\right)=n\Leftrightarrow\varepsilon_n^{-1}=p^{\frac{2n}{3+\delta}}$ for $n\in\mathbb{N}$.

    Observe that if $\mathcal{V}\in E(\varepsilon_t)$, $d\left(\mathcal{V},\sum_xp(x)\mathcal{U}_x\right)>\varepsilon_t$ for any probability distribution $p(x)$ and $\mathcal{U}_x$ whose $G$-count is less than $t(\in\mathbb{N})$.
    By using Proposition \ref{prop:probsynth}, this implies that $\{\mathcal{U}:\C(\mathcal{U},G,0)<t\}$ is not an $\sqrt{\varepsilon_t}$-covering of the $2\sqrt{\varepsilon_t}$-ball centered at $\mathcal{V}$.
    Let $\{\mathcal{V}_x\}_{x\in X}$ be a $(c_1\sqrt{\varepsilon_t})$-covering of the $2\sqrt{\varepsilon_t}$-ball centered at the identity channel with a constant $c_1\in(0,1)$. We can assume that the size of $\{\mathcal{V}_x\}_{x\in X}$ is upper bounded by a constant independent of $\varepsilon_t$ as shown in the construction of a probabilistic synthesis algorithm~\cite{akibue2024probabilistic}.
    Since $\{\mathcal{U}:\C(\mathcal{U},G,0)<t\}$ is not an $\sqrt{\varepsilon_t}$-covering, we find
    \begin{equation}
        \exists x\in X,\forall\mathcal{U}\ {\rm s.t.}\ \C(\mathcal{U},G,0)<t,d(\mathcal{V}_x\circ\mathcal{V},\mathcal{U})>c_2\sqrt{\varepsilon_t},
    \end{equation}
    where $c_2=1-c_1$.
    
    When $G=T$, by using Eq.~\eqref{eq:Tcount_IntegerPoint}, we obtain
    \begin{equation}
        \exists x\in X,\forall(\alpha,\beta,\gamma,\delta)\in S_{\mathbb{Z}[\sqrt{2}]}(2^{\frac{t}{2}-1}),d(\mathcal{V}_x\circ\mathcal{V},\mathcal{U}(\alpha,\beta,\gamma,\delta))>c_2\sqrt{\varepsilon_t}.
    \end{equation}
    Thus, Proposition \ref{prop:uncovered_area} implies
    \begin{eqnarray}
        \mu(E(\varepsilon_t))&\leq&
        \mu\left(\cup_{x\in X}\{\mathcal{V}:\forall(\alpha,\beta,\gamma,\delta)\in S_{\mathbb{Z}[\sqrt{2}]}(2^{\frac{t}{2}-1}),d(\mathcal{V}_x\circ\mathcal{V},\mathcal{U}(\alpha,\beta,\gamma,\delta))>c_2\sqrt{\varepsilon_t}\}\right)\\
        &\leq&\sum_{x\in X}\mu\left(\{\mathcal{V}:\forall(\alpha,\beta,\gamma,\delta)\in S_{\mathbb{Z}[\sqrt{2}]}(2^{\frac{t}{2}-1}),d(\mathcal{V}_x\circ\mathcal{V},\mathcal{U}(\alpha,\beta,\gamma,\delta))>c_2\sqrt{\varepsilon_t}\}\right)\\
        &=&|X|\mu\left(\{\mathcal{V}:\forall(\alpha,\beta,\gamma,\delta)\in S_{\mathbb{Z}[\sqrt{2}]}(2^{\frac{t}{2}-1}),d(\mathcal{V},\mathcal{U}(\alpha,\beta,\gamma,\delta))>c_2\sqrt{\varepsilon_t}\}\right)\\
        &\leq& c\frac{t^2}{2^t\sqrt{\varepsilon_t}^{3}}
    \end{eqnarray}
    with some positive number $c>0$,
    where we used the unitary invariance of the Haar measure to derive the equation.
    
    When $G=V_p$, by using Eq.~\eqref{eq:Vcount_IntegerPoint}, we obtain
    \begin{equation}
        \exists x\in X,\forall(\alpha,\beta,\gamma,\delta)\in S_{\mathbb{Z}}(p^{t-1}),d(\mathcal{V}_x\circ\mathcal{V},\mathcal{U}(\alpha,\beta,\gamma,\delta))>c_2\sqrt{\varepsilon_t}.
    \end{equation}
    Thus, Proposition \ref{prop:uncovered_area} implies
    \begin{equation}
        \mu(E(\varepsilon_t))\leq c\frac{t^2}{p^t\sqrt{\varepsilon_t}^{3}}
    \end{equation}
    with some positive number $c>0$.

    Hence, in both cases, we find
    $\mu(E(\varepsilon_t))\leq c\frac{t^2}{p^t\sqrt{\varepsilon_t}^3}=ct^2\left(p^{-\frac{\delta}{3+\delta}}\right)^t\leq cr^t$ with some constant $c>0$ and $r\in(0,1)$. 
    Therefore, we obtain
    \begin{equation}                        \mu\left(\bigcap_{n\in\mathbb{N}}\bigcup_{\varepsilon\in\left(0,\varepsilon_n\right)}E\left(\varepsilon\right)\right)
        \leq
        \mu\left(\bigcup_{\varepsilon\in\left(0,\varepsilon_n\right)}E\left(\varepsilon\right)\right)
        \leq
        \mu\left(\bigcup_{t\geq n}E\left(\varepsilon_t\right)\right)
        \leq\frac{c}{1-r}r^n
    \end{equation} 
    for any $n\in\mathbb{N}$, where we used $\bigcup_{\varepsilon\in\left(0,\varepsilon_n\right)}E\left(\varepsilon\right)\subseteq\cup_{t\geq n}E(\varepsilon_t)$ to derive the second inequality. This completes the proof.    
\end{proof}

\begin{proof}[Proof of Theorem \ref{thm: G-rate equal 3}]
    Lemma \ref{lemma:aelower} and Lemma \ref{lemma:aeupper} imply $\CO(\mathcal{U},G)=3$ a.e..
    Combining with Lemma \ref{lemma:aeupper_prob} and Proposition \ref{prop:Gcountprobdet}, we obtain $\CO^{\prob}(\mathcal{U},G)=\frac{3}{2}$ a.e..    
\end{proof}

\begin{theorem}\label{thm: G-lower geq 4}
Let $G$ be either $T$ or $V_p$ with an odd prime $p$.
    Let $U=\begin{pmatrix}
        a+ib&-c+id\\
        c+id&a-ib
    \end{pmatrix}$
    induce a unitary channel $\mathcal{U}$. Assume that $\mathcal{U}$ is not exactly synthesizable by Clifford+$G$ gates. 
    \begin{enumerate}
        \item If $G=T$ and $a:b:c:d$ can be represented by $\mathbb{Z}[\sqrt{2}]$, we have
        \[
            2\underline{\CO}^{\prob}(\mathcal{U},G)\geq\underline{\CO}(\mathcal{U},G)\geq4.
        \]
        \item If $G=V_p$ and $a:b:c:d$ can be represented by $\mathbb{Z}$, we have
        \[
            2\underline{\CO}^{\prob}(\mathcal{U},G)\geq\underline{\CO}(\mathcal{U},G)\geq4.
        \]
    \end{enumerate}
\end{theorem}
Due to Proposition \ref{prop:Gcountprobdet}, it is sufficient to prove the statements for $\underline{\CO}(\mathcal{U}, G)$.
Since the proof relies on advanced results from Diophantine approximation, it is deferred to the next section.

\begin{theorem}\label{thm: indeterminate G-rate}
Let $G$ be either $T$ or $V_p$ with an odd prime $p$.    There exist unitary channels whose exact G-count order is not defined.
\end{theorem}
To prove this theorem, we use the following proposition.
\begin{proposition}\label{prop: approximationspecial}
    For all $c > 1$, $C_1 > 0$, $C_2 > 0$ and a number $A \in \bbC$ algebraic over $\bbQ$, we have
    \begin{equation}\label{eq: conj1}
        \#\left\{
        \frac{x}{\sqrt{2}^k}\ \middle|\ k \in \Z_{\geq 0},\ x\in \Z[\sqrt{2}],\ 
        \left|\frac{x}{\sqrt{2}^k} - A\right|<\frac{C_1}{2^{ck}}\text{ and }
        \left|\frac{x^\bullet}{\sqrt{2}^k}\right|\leq C_2\right\}
        < \infty,
    \end{equation}
    where $x^\bullet$ represents the Galois conjugate, defined as $(a+b\sqrt{2})^\bullet=a-b\sqrt{2}$ for $a,b\in\mathbb{Z}$.
\end{proposition}
\begin{proof}
    Setting $S = M_{\bbQ(\sqrt{2})}^{\infty}\cup\{\sqrt{2}\bbZ[\sqrt{2}]\}$ and $K=\bbQ(\sqrt{2})$ in Proposition \ref{prop: approximation}, which is shown in the next section, completes the proof.
\end{proof}

\begin{proof}[Proof of Theorem~\ref{thm: indeterminate G-rate}.]
    While we provide an example using Clifford+$T$, extending it to generalized $V$ gates is straightforward.

    Let $\{\mathcal{U}_n\}_{n\in\mathbb{N}}$ be a set of Clifford+$T$ unitary channels such that
    \begin{equation}
        \frac{1}{2}\varepsilon_n\leq d(\mathcal{U}_n,id)\leq\varepsilon_n,\quad 
        \C(\mathcal{U}_n,T,0)\leq c'\log_2\left(\frac{1}{\varepsilon_n}\right),
    \end{equation}
    where $\varepsilon_n^{-1}=2^{n!}$ and $c'>0$ is a constant. Define $\mathcal{V}_m:=\mathcal{U}_1\circ\mathcal{U}_2\circ\cdots\circ\mathcal{U}_m$, $\mathcal{U}:=\lim_{n\to\infty}\mathcal{V}_n$ and $\eta_m:=\sum_{n=m}^\infty\varepsilon_n$. Let $t\in\mathbb{R}$ be $t<\underline{\CO}(\mathcal{U},T)$. Since
    \begin{equation}
        \C(\mathcal{U},T,\varepsilon)\leq\sum_{n=1}^m\C(\mathcal{U}_n,T,0)
        \ \ \ {\rm if}\ \varepsilon\geq \sum_{n=m+1}^\infty\varepsilon_n,
    \end{equation}
    there exists $M\in\mathbb{R}$ such that for any $m\geq M$, it holds
    \begin{align}
        &\phantom{=} t((m+1)!)-t< 
        t\log_2\left(\frac{1}{\sum_{n=m+1}^\infty\varepsilon_n}\right)\leq    \C\left(\mathcal{U},T,\sum_{n=m+1}^\infty\varepsilon_n\right)\\
        &\leq \sum_{n=1}^m\C(\mathcal{U}_n,T,0)
        \leq c'\sum_{n=1}^m\log_2\left(\frac{1}{\varepsilon_n}\right)=c'\sum_{n=1}^mn!\leq2c'(m!),
    \end{align}
    where in the first inequality, we used the following calculation
    \begin{equation}
        \sum_{n=m+1}^\infty\varepsilon_n=\sum_{n=m+1}^\infty\frac{1}{2^{n!}}< \frac{1}{2^{(m+1)!}}+\frac{1}{2^{(m+1)!+1}}+\frac{1}{2^{(m+1)!+2}}\cdots
        =\frac{2}{2^{(m+1)!}}
    \end{equation}
    and in the last inequality, we used the following calculation
    \begin{align}
        &\phantom{=} \sum_{n=1}^mn!=m!\left(1+\frac{1}{m}+\frac{1}{m(m-1)}+\frac{1}{m(m-1)(m-2)}+\cdots+\frac{1}{m!}\right)\\
        &\leq m!\left(1+\frac{1}{m}+\frac{1}{m(m-1)}+\frac{1}{(m-1)(m-2)}+\cdots+\frac{1}{2\cdot1}\right)=2(m!).
    \end{align}
    This implies that $t\leq0$ since $\lim_{m\to\infty}\frac{m!}{(m+1)!-1}=0$. Thus, $\underline{\CO}(\mathcal{U},T)=0$.

To show $\overline{\CO}(\mathcal{U},T)\geq4$, we first show that for any $c>1$, there exists $\varepsilon_0>0$ such that the inequality $\C(\mathcal{V},T,0)\geq\frac{4}{c}\log_2\left(\frac{1}{\varepsilon}\right)$ holds for any $\varepsilon\in(0,\varepsilon_0)$ and any Clifford+$T$ unitary channel $\mathcal{V}$ satisfying $d(\mathcal{V},id)\in(0,\varepsilon]$ by using Proposition \ref{prop: approximationspecial}.
As shown by Kliuchnikov et al.~\cite{kliuchnikov2013fast}, the unitary operator $V$ associated with $\mathcal{V}$ can take one of the following two possible forms.

\begin{enumerate}
    \item Suppose $V=\frac{1}{\sqrt{2}^k}
        \begin{pmatrix}
            \alpha+i\beta&-\gamma+i\delta\\
            \gamma+i\delta&\alpha-i\beta
        \end{pmatrix}$ ($\alpha,\beta,\gamma,\delta\in\mathbb{Z}[\sqrt{2}]$). The inequality $d(\mathcal{V},id)\in(0,\varepsilon]$ implies $\frac{\alpha}{\sqrt{2}^k}\in[\sqrt{1-\varepsilon^2},1)$ by using Eq.~\eqref{eq:diamond_distance_1Q}. Thus, $0<\left|\frac{\alpha}{\sqrt{2}^k}-1\right|<\varepsilon^2$.
    Since $V^\bullet\in \SU(2)$, we obtain $\frac{|\alpha^\bullet|}{\sqrt{2}^k}\leq 1$, where $V^\bullet$ denotes the matrix whose elements are the Galois conjugate of those of $V$.

    \item Suppose $V=\frac{1}{\sqrt{2}^k}\begin{pmatrix}
        \alpha+i\beta&-\gamma+i\delta\\
        \gamma+i\delta&\alpha-i\beta
    \end{pmatrix}R_z\left(\frac{\pi}{4}\right)$ ($\alpha,\beta,\gamma,\delta\in\mathbb{Z}[\sqrt{2}]$). By using Eq.~\eqref{eq:diamond_distance_1Q}, the inequality $d(\mathcal{V},id)\in(0,\varepsilon]$ implies 
    \begin{equation}
        \begin{pmatrix}
        \cos\frac{\pi}{8}\\\sin\frac{\pi}{8}
    \end{pmatrix}\cdot\begin{pmatrix}
        \alpha\\\beta
    \end{pmatrix}=\cos\frac{\pi}{8}(\alpha+(\sqrt{2}-1)\beta)\in\sqrt{2}^k[\sqrt{1-\varepsilon^2},1).
    \end{equation}
    Thus, we have
    \[
        0<\left|\frac{\alpha+(\sqrt{2}-1)\beta}{\sqrt{2}^k}-\sqrt{4-2\sqrt{2}}\right|<\varepsilon^2.
    \]
     Since $(VR_z(-\frac{\pi}{4}))^\bullet$ is in $\SU(2)$, we obtain
     \begin{equation}
         \frac{|(\alpha+(\sqrt{2}-1)\beta)^\bullet|}{\sqrt{2}^{k}}\leq\frac{1}{\sqrt{2}^{k}}\lpnorm{2}{\begin{pmatrix}
             \alpha\\\beta
         \end{pmatrix}^\bullet}
         \lpnorm{2}{\begin{pmatrix}
             1\\-\sqrt{2}-1
         \end{pmatrix}}
         \leq2\sqrt{2}\cos\frac{\pi}{8}=\sqrt{4+2\sqrt{2}}.
     \end{equation}
\end{enumerate}
In both cases, Proposition \ref{prop: approximationspecial} implies that
for all $c>1$, there exists $k_0>0$ such that $\forall k\geq k_0,\varepsilon^2>\frac{1}{2^{ck}}$ if $d(\mathcal{V},id)\in(0,\varepsilon]$.
Otherwise, there are infinitely many $\frac{\alpha}{\sqrt{2}^k}$ satisfying $\left|\frac{\alpha}{\sqrt{2}^k}-1\right|<\frac{1}{2^{ck}}$ and $\left|\frac{\alpha^\bullet}{\sqrt{2}^k}\right|\leq1$, or
infinitely many $\frac{\alpha+(\sqrt{2}-1)\beta}{\sqrt{2}^k}$ satisfying
$\left|\frac{\alpha+(\sqrt{2}-1)\beta}{\sqrt{2}^k}-\sqrt{4-2\sqrt{2}}\right|<\frac{1}{2^{ck}}$ and $\frac{|(\alpha+(\sqrt{2}-1)\beta)^\bullet|}{\sqrt{2}^{k}}\leq\sqrt{4+2\sqrt{2}}$, which contradicts Proposition \ref{prop: approximationspecial}.

Since we can assume $k\leq\frac{1}{2}(\C(\mathcal{V},T,0)+5)$~\cite{giles2019remarksmatsumotoamanosnormal}, we obtain that there exists $\varepsilon_0>0$ such that $\C(\mathcal{V},T,0)> \frac{4}{c}\log_2\left(\frac{1}{\varepsilon}\right)-5$ for any $\varepsilon\in(0,\varepsilon_0)$ and any Clifford+T unitary channel $\mathcal{V}$ satisfying $d(\mathcal{V},id)\in(0,\varepsilon]$.

This implies that
\begin{align}
    \C(\mathcal{U},T,\varepsilon)
    &\geq \C(\mathcal{U}_{m+1}\circ\mathcal{U}_{m+2}\circ\cdots,T,\varepsilon)-\C(\mathcal{V}_m^{-1},T,0)\\
    &= \C(\mathcal{U}_{m+1}\circ\mathcal{U}_{m+2}\circ\cdots,T,\varepsilon)-\C(\mathcal{V}_m,T,0)\\
    &\geq \frac{4}{c}\log_2\left(\frac{1}{\varepsilon+\sum_{n=m+1}^\infty\varepsilon_n}\right)-c'\sum_{n=1}^m\log_2\left(\frac{1}{\varepsilon_n}\right)
\end{align}
where we assume assume $\varepsilon=\frac{1}{4}\varepsilon_{m+1}$, $m$ is large enough to satisfy $\frac{9}{4}\varepsilon_{m+1}<\varepsilon_0$, and we use 
\begin{align}
   d(\mathcal{U}_{m+1}\circ\mathcal{U}_{m+2}\circ\cdots, id)
   &\geq d(\mathcal{U}_{m+1},id)-d(\mathcal{U}_{m+2}\circ\mathcal{U}_{m+3}\circ\cdots, id)\\
   &\geq \frac{1}{2}\varepsilon_{m+1}-\sum_{n=m+2}^\infty\varepsilon_n
   > \frac{1}{2}\varepsilon_{m+1}-\frac{2}{2^{(m+2)!}}\geq\frac{1}{4}\varepsilon_{m+1}
\end{align}
in the second inequality.

By for any $t>\overline{\CO}(\mathcal{U},T)$, 
we obtain that there exists $M$ such that for any $m\geq M$, it holds that
\begin{align}
    t((m+1)!+2) &\geq \frac{4}{c}\log_2\left(\frac{1}{\frac{1}{4}\varepsilon_{m+1}+\sum_{n=m+1}^\infty\varepsilon_n}\right)-c'\sum_{n=1}^m\log_2\left(\frac{1}{\varepsilon_n}\right)\\
    &\geq \frac{4}{c}(m+1)!-\frac{4}{c}\log_2\left(\frac{9}{4}\right)-2c'(m!).
\end{align}
Since $\lim_{m\to\infty}\frac{(RHS)}{(m+1)!+2}=\frac{4}{c}$, we obtain $t\geq\frac{4}{c}$. Since this holds for any $c>1$, we obtain $\overline{\CO}(\mathcal{U},T)\geq4$.

\end{proof}
Since this construction is very similar to that of a Liouville number, we refer to such unitary channels as Liouville-type.

\section{Big hole for arithmetic gates}

We prove the following general theorem to derive Theorem \ref{thm: G-lower geq 4}.
\begin{theorem}\label{thm: ldhlower geq 2}
    Let $K$ be a totally real number field, $S$ be a finite subset of $M_K$ containing $M_K^{\infty}$, and $\calX$ be a finite subset of $\overline{\bbQ}$.
    Let $\scrC$ be a subset of $\AQ(K,S,\calX)$.
    Let $U=\begin{pmatrix}
        a+ib&-c+id\\
        c+id&a-ib
    \end{pmatrix}$
    realize a unitary transformation $\calU$.
    If $\calU$ is not realized by any elements in $\scrC$, and $a:b:c:d$ can be represented by $\calO_K$, we have $\ldhlower_{\scrC}(U) \geq 2$.
\end{theorem}

The following Diophantine approximation result is the essential part of the proof of Theorem \ref{thm: ldhlower geq 2}.

\begin{proposition}\label{prop: approximation}
    Let $K$ be a totally real Galois extension of $\bbQ$, and
    $S$ be a finite subset of $M_K$ containing $M_K^{\infty}$.
    Let $\sigma_1, \sigma_2, \ldots, \sigma_{[K:\bbQ]}$ be the all embeddings of $K$ into $\bbR$.
    Extend each normalized absolute value $\|\cdot\|_{\sigma_i}$ to the algebraic closure $\overline{\bbQ}$ in one way and denote it by the same notation.
    For $c > 1$, and $C_i > 0\ (1\leq i \leq [K:\bbQ])$ and $A \in \overline{\bbQ}$, the set
    \begin{equation}\label{eq: conj2}
        \scrA = \left\{
        \frac{x}{u}\ \middle|\ 
        \begin{array}{l}
            x,u\in \calO_K, \Supp(u)\subset S,
            \left\|\frac{x}{u} - A\right\|_{\sigma_1}<\frac{C_1}{H_K(u)^{c}}\text{ and }\\
            \left\|\frac{x}{u}\right\|_{\sigma_i}\leq C_i\ (2\leq i \leq [K:\bbQ])
        \end{array}
        \right\}
    \end{equation}
    is finite.
\end{proposition}

The contents of this paper can be separated into two parts.
The first part is Section \ref{sec: Diophantine approximations}.
In this section, we present purely number-theoretic results.
The second part is Section \ref{sec: Notation} and Section \ref{sec: Proof of main theorem}.
In these sections, we deal with the estimation of the asymptotic $G$-count.

Section \ref{subsec: absolute values} is devoted to recalling some notation and fundamental results on the absolute values and height functions.
In Section \ref{subsec: subspace theorem}, we recall a powerful Diophantine approximation result.
We will prove Proposition \ref{prop: approximation} in Section \ref{subsec: proof of approximation}.

In Section \ref{subsec: Clifford+T} and Section \ref{subsec: Clifford+V}, we will recall the notation used in Theorem \ref{thm: G-lower geq 4}.
In Section \ref{subsec: arithmetic quantum}, we introduce the notion of arithmetic quantum matrices and the least denominator height, which can be regarded as a generalization of the $T$-count and $V$-count.
We explain how Theorem \ref{thm: ldhlower geq 2} implies Theorem \ref{thm: G-lower geq 4} in Section \ref{subsec: how to use}.
The proof of Theorem \ref{thm: ldhlower geq 2} is given in Section \ref{sec: Proof of main theorem}.


\subsection{Diophantine approximations}\label{sec: Diophantine approximations}

    In Section \ref{subsec: absolute values}, we recall the definition and some basic facts on absolute values on number fields.
    In Section \ref{subsec: subspace theorem}, we describe a subspace theorem, one of the most powerful Diophantine approximation results.
    We will prove Proposition \ref{prop: approximation} in Section \ref{subsec: proof of approximation} using the subspace theorem.

\subsubsection{Preparation of absolute values}\label{subsec: absolute values}
   
    \begin{definition}[Absolute values]\label{dfn: absolute values}
        Let $K$ be a field.
        A map $|\cdot|_v \colon K \longrightarrow \bbR$ is called an {\it absolute value} if the following conditions \ref{item: dfn: positivity}-\ref{item: dfn: triangle inequality} hold.
        \begin{enumerate}
            \item\label{item: dfn: positivity} $|a|_v\geq 0$ for all $a\in K$,
            \item $|a|_v = 0$ if and only if $a=0$,
            \item $|ab|_v = |a|_v |b|_v$ for all $a,b \in K$, and
            \item\label{item: dfn: triangle inequality} $|a+b|_v \leq |a|_v + |b|_v$ for all $a,b\in K$.
        \end{enumerate}
        If $|\cdot|_v$ satisfies the following stronger condition (iv)' than \ref{item: dfn: triangle inequality}, it is said to be {\it non-Archimedean}.
        \begin{enumerate}
            \item[(iv)']$|a+b|_v \leq \max\{|a|_v, |b|_v\}$ for all $a,b\in K$.
        \end{enumerate}
        When an absolute value $|\cdot|_v$ is not a non-Archimedean absolute value, it is said to be {\it Archimedean}.

        For an absolute value $|\cdot|_v\colon K\longrightarrow \R$, the function $d_v\colon K\times K \longrightarrow \bbR$ defined by $d_v(x,y) = |x-y|_v$ is a distance function.
        The distance function $d_v$ induces a topology on $K$.
        When two absolute values $|\cdot|_{v_1}$ and $|\cdot|_{v_2}$ induces the same topology on $K$, we say that $|\cdot|_{v_1}$ and $|\cdot|_{v_2}$ are {\it equivalent}.
    \end{definition}
    \begin{example}[Absolute values on number field]
        We give some important examples of absolute values.
        \begin{enumerate}
            \item\label{item: ex: trivial absolute value}
                For any field $K$, the map $|\cdot|_\text{triv}$ defined by
                \[
                    |a|_{\text{triv}} =
                    \begin{cases}
                        0 & \text{if }a=0,\\
                        1 & \text{otherwise}
                    \end{cases}
                \]
                is called the {\it trivial absolute value}.
            
            \item\label{item: ex: standard absolute values on R, C}
                The standard absolute values $|\cdot| \colon \R\longrightarrow \R$ defined by $|a| \coloneqq \max\{ a, -a\}$ and $|\cdot| \colon \bbC \longrightarrow \R$ defined by $|a+b\sqrt{-1}|\coloneqq \sqrt{a^2+b^2}$ are of course absolute values on $\bbR$ and $\bbC$, respectively.

            \item\label{item: ex: infinite places of Q}
                The restriction of standard absolute value $|\cdot|$ to $\Q$ is written by $|\cdot|_\infty$.
            
            \item\label{item: ex: finite places of Q}
                For a prime number $p\in \Z$ and a non-zero integer $a$, define
                \[
                    \ord_p(a) = \max\{n\in \Z \mid n\geq 0,\ p^n \text{ devides }a\}.
                \]
                The map $|\cdot|_p \colon \Q \longrightarrow \R$ defined by $|0|_p \coloneqq 0$ and
                \[
                    \left| \frac{a}{b} \right|_p \coloneqq p^{\ord_p(b)-\ord_p(a)} \quad \text{for } a, b \in \Z\setminus\{0\}
                \]
                is called the {\it $p$-adic absolute value}. It is easy to see that $|\cdot|_p$ is actually an absolute value on $\Q$.

            \item\label{item: ex: infinite places of number field}
                Let $K$ be a number field.
                For a field embedding $\sigma \colon K\longrightarrow \bbC$, the map $|\cdot|_\sigma \colon K \longrightarrow \bbR$ defined by $|a|_{\sigma} \coloneqq |\sigma(a)|$ is an absolute value.
                Note that two embeddings $\sigma$ and $\iota \circ \sigma$ define the same absolute value, where $\iota$ is the complex conjugate. This absolute value is a generalization of the standard absolute value $|\cdot|_\infty$ on $\Q$ to general number fields.
                
            \item\label{item: ex: finite places of number field}
                Let $K$ be a number field, $\calO_K$ the ring of integers of $K$, and $\frakp$ a non-zero prime ideal of $\calO_K$.
                For an element $a\in \calO_K\setminus\{0\}$, define the order at $\frakp$ by
                \[
                    \ord_\frakp(a) \coloneqq \max\{ n \in \Z \mid n\geq 0,\ a \in \frakp^n \}.
                \]
                The ideal $\frakp \cap \Z$ is generated by a prime number $p$.
                In this situation, the ramification degree $e(\frakp/p)$ is defined by $\ord_{\frakp}(p)$.
                The map $|\cdot|_{\frakp} \colon K \longrightarrow \R$ defined by $|0|_\frakp \coloneqq 0$ and
                \[
                    \left| \frac{a}{b} \right|_{\frakp} \coloneqq p^{(\ord_{\frakp}(b)-\ord_{\frakp}(a))/e(\frakp/p)} \quad \text{for } a, b \in \calO_K \setminus\{0\}
                \]
                is called the {\it $\frakp$-adic absolute value}. It is easy to see that $|\cdot|_\frakp$ is an absolute value on $K$. The restriction of $|\cdot|_\frakp$ to $\Q$ coincides with $|\cdot|_p$.
        \end{enumerate}
    \end{example}

    \begin{theorem}[{Minkowski's Theorem, \cite[Theorem ?]{Neu}}]
        For a number field $K$, any absolute value $|\cdot|_v$ on $K$ is equivalent to either
        \begin{itemize}
            \item $|\cdot|_{\text{\rm triv}}$,
            \item $|\cdot|_\sigma$ for some field embedding $\sigma \colon K \longrightarrow \bbC$, or
            \item $|\cdot|_{\frakp}$ for some non-zero prime ideal $\frakp$ of $\calO_K$. 
        \end{itemize}
    \end{theorem}
    
    \begin{definition}[the naive height function on the projective space]\label{dfn: height function}
        Let $K$ be a number field. Set
        \[
            M_K \coloneqq \{ |\cdot|_v \mid v \text{ is a field embedding }\sigma \colon K\longrightarrow \bbC \text{ or a non-zero prime ideal }\frakp \text{ of }\calO_K\}.
        \]
        Let $M_K^{\infty}$ be the set of Archimedean absolute values in $M_K$.
        We simply refer $v$ for $|\cdot|_v$.
        For $v\in M_K$, set $K_v$ and $\Q_v$ to be the completion of $K$ and $\Q$, respectively, with respect to the distance function $d_v$ defined in Definition \ref{dfn: absolute values}.
        Let $n_v$ be the extension degree $[K_v:\Q_v]$.
        We simply write $|\cdot|_v^{n_v}$ as $\|\cdot\|_v$.
        
        Then, the {\it relative multiplicative height} $H_{K,\bbP^n} \colon \bbP^n(K) \longrightarrow \R$ is defined by
        \[
            H_{K,\bbP^n}([x_0,x_1,\ldots,x_n]) \coloneqq \prod_{v \in M_K} \max\{ \|x_i\|_v\mid 0\leq i \leq n\}
        \]
        for $(x_0,x_1,\ldots, x_n) \in K^{n+1}\setminus\{(0,0,\ldots, 0)\}$.
        We must note that $H_{K,\bbP^n}$ depends on the base field $K$. This fact motivates the following definition (see also Remark \ref{rem: remarks on heights}\ref{item: rem: independence of base fields}).
        Regarding $x\in K$ as an element $[1,x] \in \bbP^1(K)$, we define the functions $H_K \colon K \longrightarrow \R$.
    \end{definition}

    \begin{remark}\label{rem: remarks on heights}
        We give some remarks on the definition of the heights.
        \begin{enumerate}
            \item \label{item: rem: independence of expression of points}
                Since we have the product formula
                \begin{equation}\label{eq: product formula}
                    \prod_{v\in M_K} \|a\|_v = 1
                \end{equation}
                for all $a\in K\setminus\{0\}$, the definition of $H_{\bbP^n,K}$ does not depend on the expression of the point $P\in \bbP^n(K)$.
            \item \label{item: rem: independence of base fields}
                Note that $H_{K,\bbP^n}$ depends on the base field $K$.
                For example, we have the following equalities for $K=\Q(\sqrt{2})$ and $n=1$:
                \begin{align}
                    H_{K,\bbP^1}([1,2]) &= 4, \text{ and}\\
                    H_{\Q, \bbP^1}([1,2]) &= 2.
                \end{align}
                In general, for a field extension $K'/K$ and an element $P = [x_0,x_1,\ldots, x_n]\in \bbP^n(K)$, we have
                \[
                    H_{K', \bbP^n}(P) = H_{K,\bbP^n}(P)^{[K':K]}.
                \]
                This equality implies the independence of $H_{\bbP^n}$ on the base number field $K$.
        \end{enumerate}
    \end{remark}

    The following theorem, known as Northcott's finiteness property, is used to prove the finiteness of rational points with some properties in Diophantine Geometry. See \cite{HS} for more general statements.
    \begin{theorem}\label{thm: Northcott property}
        For any number field $K$ and any constant $B\in \R$, the set
        \[
            \left\{ P\in \bbP^1(K) \mid H_K(P) \leq B \right\}
        \]
        is a finite set.
    \end{theorem}

    \begin{example}\label{ex: absolute values}
        For \( K = \mathbb{Q}(\sqrt{2}) \), we present some calculations that will be used in the proof of the main result. There are only two embeddings of \( K \) into \( \mathbb{C} \). These embeddings map rational numbers identically and send \( \sqrt{2} \) to either \( \sqrt{2} \) or \( -\sqrt{2} \). Let \( \sigma_1 \) be the embedding such that \( \sigma_1(\sqrt{2}) = \sqrt{2} \), and let \( \sigma_2 \) be the other embedding. The images of both \( \sigma_1 \) and \( \sigma_2 \) are contained in \( \mathbb{R} \). Therefore, we have \( n_{\sigma_i} = [\mathbb{R}:\mathbb{R}] = 1 \) for \( i = 1, 2 \).

        There is only one non-Archimedean absolute value $|\cdot|_{\frakp}$ whose restriction to $\bbQ$ coincide with $|\cdot|_2$.
        The corresponding prime ideal
        \[
            \frakp = \{ a\in K \ |\ |a|_{\frakp} < 1\}
        \]
        is generated by $\sqrt{2}$, and we have $n_\frakp = [K_\frakp : \bbQ_2] = [\bbQ_2(\sqrt{2}): \bbQ_2] = 2$.
    \end{example}

\subsubsection{Subspace theorem}\label{subsec: subspace theorem}
    \begin{theorem}[{\cite{Schm80} for $K=\bbQ$ and \cite{Schl77} in general}]\label{thm: subspace theorem}
        Let $K$ be a number field with a ring of integers $\calO_K$.
        Let $S$ be a finite subset of $M_K$ and extend $v \in S$ to $\overline{\bbQ}$ in one way.
        For each $v \in S$, let $L_{v,i}$ $(0\leq i \leq n)$ be $n+1$ linearly independent linear forms in $n+1$ variables, with coefficients in $\overline{\bbQ}$.
        For a tuple $s = (s_0, s_1, \ldots, s_n) \in \calO_K^{n+1}$, define the size of $s$ as
        \[
            \mathrm{size}(s) \coloneqq \max\left\{ \|s_i\|_v\ |\ v \in M_K^{\infty}, 0\leq i \leq n\right\}.
        \]
        Fix $\varepsilon > 0$. Let $Q$ be the set of all $s\in \calO_K^{n+1}$ satisfying the inequality
        \[
            \prod_{v\in S} \prod_{i=0}^n \|L_{v,i}(s)\|_v < \mathrm{size}(s)^{-\varepsilon}.
        \]
        Then, $Q$ is contained in a finite union of proper linear subspaces of $\overline{\bbQ}^{n+1}$.
    \end{theorem}
    
    Although the following Thue-Siegel-Roth's theorem is not used in the proof of the main theorem, it should be remarked upon to explain the strength of Theorem \ref{thm: subspace theorem}.
    
    \begin{theorem}[{\cite{Rot55}}]\label{thm: Roth}
        For any real algebraic number $\alpha$ and any positive real number $\varepsilon > 0$, the set of $p/q \in \bbQ$ with
        \[
            \left| \alpha - \frac{p}{q}\right| < \frac{1}{q^{2+\varepsilon}}
        \]
        is finite.
    \end{theorem}
    Setting $K= \bbQ$, $n = 1$, $S = \{|\cdot|_\infty\}$,
    \begin{align}
        L_0(x_0, x_1) &= x_0, \text{ and}\\
        L_1(x_0, x_1) &= \alpha x_0 - x_1
    \end{align}
    with an algebraic real number $\alpha$,
    one can see that Theorem \ref{thm: subspace theorem} implies Theorem \ref{thm: Roth}.

\subsubsection{Proof of Proposition \ref{prop: approximation}}\label{subsec: proof of approximation}

In this subsection, we prove Proposition \ref{prop: approximation}.
Initially, we present a preparation lemma.
\begin{lemma}\label{lem: preparation}
    Let notation as in Proposition \ref{prop: approximation}.
    Then, the set
    \[
        \scrB = \left\{ \frac{x}{u} \in \scrA\ \middle|\ \mathrm{size}(x,u) > H_K(u)^2 \left(\prod_{i=1}^{[K:\bbQ]} C_i\right)^{2/(c-1)}\right\}
    \]
    is finite.
\end{lemma}
\begin{proof}
    To ease the notion, let $C_0 = \prod_{i=1}^{[K:\bbQ]} C_i$ and $C_0' = \max\left(\{1\}\cup\{C_i\ |\ 1\leq i\leq [K:\bbQ]\}\right)$.
    For $x/u \in \scrB$, we have the inequalities
    \begin{align}
        \mathrm{size}(x,u)
        &= \max\{ \|x\|_{\sigma_i}, \|u\|_{\sigma_i} \ |\ 1\leq i \leq [K:\bbQ]\}\\
        &<\max\{ \max\{C_i,1\}\|u\|_{\sigma_i} \ |\ 1\leq i \leq [K:\bbQ]\}\\
        &\leq C_0'\max\{ \|u\|_{\sigma_i} \ |\ 1\leq i \leq [K:\bbQ]\}
        \leq C_0' H_K(u),
    \end{align}
    and
    \begin{align}
        \mathrm{size}(x,u) > C_0 H_K(u)^2.
    \end{align}
    Combining them, we get
    \begin{equation}
        H_K(u) < C_0^{-1} C_0'.
    \end{equation}
    By Theorem \ref{thm: Northcott property}, the number of such $u$ is at most finite.
    For each such $u$, since we have
    \begin{equation}
        H_K(x) = \prod_{i=1}^{[K:\bbQ]} \max\{1, \|x\|_{\sigma_i}\}
        \leq \left(\| A \|_{\sigma_1} + \frac{C_1}{H_K(u)^c}\right)\|u\|_{\sigma_1} \prod_{i=2}^{[K:\bbQ]} C_i \|u\|_{\sigma_i},
    \end{equation}
    the number of such $x$ is at most finite again by Theorem \ref{thm: Northcott property}.
    Thus, the set $\scrB$ is finite.
\end{proof}

\begin{proof}[Proof of Proposition \ref{prop: approximation}]
    Let $C_0, C_0'$ be as in the proof of Lemma \ref{lem: preparation}.
    By Lemma \ref{lem: preparation}, it is enough to show that the set
    \[
        \scrA' \coloneqq \left\{ \frac{x}{u} \in \scrA\ \middle|\ \mathrm{size}(x,u) \leq H_K(u)^2 C_0^{2/(c-1)}\right\}
    \]
    is finite.
    Consider the following linear forms
    \[
    \begin{array}{lll}
        L_{\sigma_1,0}(x_0,x_1) \coloneqq x_0 - Ax_1,
        &L_{\sigma_1,1}(x_0,x_1) \coloneqq x_1,\\
        L_{v,0}(x_0,x_1) \coloneqq x_0,
        &L_{v,1}(x_0,x_1) \coloneqq x_1 &(\text{for }v\in S\setminus\{\sigma_1\}).
    \end{array}
    \]
    For $x/u \in \scrA'$, we have the inequalities
    \begin{align}
        &\hphantom{=}\prod_{v\in S}\prod_{j=0,1} \| L_{v,j}(x,u)\|_{v}\\
        &< \frac{C_1\|u\|_{\sigma_1}}{H_K(u)^c} \cdot \|u\|_{\sigma_1} \cdot \prod_{i=2}^{[K:\bbQ]}\|x\|_{\sigma_i} \|u\|_{\sigma_i}
        \prod_{\frakp \in S \setminus M_K^{\infty}}\|x\|_{\frakp} \|u\|_{\frakp}\\
        &= \frac{C_1\|u\|_{\sigma_1}}{H_K(u)^c} \cdot \prod_{i=2}^{[K:\bbQ]}\|x\|_{\sigma_i}
        \prod_{\frakp \in S \setminus M_K^{\infty}}\|x\|_{\frakp} &\text{by Eq.~\eqref{eq: product formula} and }\Supp(u) \subset S\\
        &\leq \frac{C_1\|u\|_{\sigma_1}}{H_K(u)^c} \cdot \prod_{i=2}^{[K:\bbQ]}\|x\|_{\sigma_i}
        &\text{by }x\in \calO_K\\
        &\leq \frac{\prod_{i=1}^{[K:\bbQ]} C_i \|u\|_{\sigma_i}}{H_K(u)^c}\\
        &\leq \frac{\left(\prod_{i=1}^{[K:\bbQ]} C_i\right) \cdot H_K(u)}{H_K(u)^c}\\
        &\leq \frac{1}{\mathrm{size}(x,u)^{(c-1)/2}} &\text{by }x/u \in \scrA'.
    \end{align}
    Consequently, the number of $x/u\in \scrA'$ is finite by Theorem \ref{thm: subspace theorem}.
\end{proof}


\subsection{Arithmetic quantum matrices and its generalized counts}\label{sec: Notation}

Section \ref{subsec: Clifford+T} and Section \ref{subsec: Clifford+V} are devoted to recalling some properties of Clifford$+T$ operators and Clifford$+V$ operators, respectively.
After that, we give a definition and notation of arithmetic quantum matrices and their generalized counts in Section \ref{subsec: arithmetic quantum}.
In Section \ref{subsec: how to use}, we explain that Clifford$+T$ and Clifford$+V$ matrices are, in fact, arithmetic quantum matrices.
Moreover, we explain how Theorem \ref{thm: ldhlower geq 2} implies Theorem \ref{thm: G-lower geq 4}.
For a positive integer $n$, we write $\zeta_n$ for $\exp(2\pi i/n)$.

\subsubsection{Clifford+T matrices}\label{subsec: Clifford+T}
\begin{definition}
    {\it Clifford$+T$ matrix} is a unitary matrix given by a finite product of the following matrices
    \begin{equation}
        T=\begin{pmatrix}
        1&0\\
        0&\zeta_8
    \end{pmatrix},\quad 
    S=\begin{pmatrix}
        1&0\\
        0&i
    \end{pmatrix},\quad 
    H=\frac{1}{\sqrt{2}}\begin{pmatrix}
        1&1\\
        1&-1
    \end{pmatrix}.
    \end{equation}
\end{definition}

\begin{definition}
    For $U\in U(2)$, its $T$-count $\C(U,T,0)$ is defined as
    \begin{equation}
        \C(U,T,0):=\min\left\{N\in\mathbb{N}\ \middle|\
        \begin{array}{l}
            m\geq 1,\ d(U,g_1g_2\cdots g_m)=0 \text{ for some Clifford$+T$ gates }g_1,\ldots, g_m\\ \text{ and } \#\{i\ |\ g_i = T\} = N
        \end{array}
        \right\}.
    \end{equation}
    If the set is empty, we define $\C(U,T,0)=\infty$.
\end{definition}

\begin{definition}
    For $z \in \bbZ[\frac{1}{\sqrt{2}}, i]$ and $x \in \bbZ[\zeta_8]$, the least denominator exponent $\lde(z,x)$ of $z$ with respect to $x$ is defined by
    \[
        \lde(z,x) \coloneqq \min\{k\in \Z\ |\ zx^k \in \bbZ[\zeta_8]\}.
    \]
    If no such $k$ exists, we let $\lde(z,x) = \infty$ for convenience.
\end{definition}

\begin{proposition}[{\cite[Theorem 1]{kliuchnikov2013fast}}]\label{prop: canonical form of Clifford+T 1}
    A unitary matrix $U$ is Clifford$+T$ if and only if its entries are in the ring $\bbZ[i, 1/\sqrt{2}]$.
\end{proposition}

\begin{proposition}\label{prop: canonical form of Clifford+T 2}
    A unitary matrix $U \in U(2)$ is Clifford$+T$ if and only if $U$ is of the form
    \begin{equation}\label{eq: canonical form of Clifford+T}
        \frac{1}{\sqrt{2}^k}
        \left(\frac{(1+\sqrt{2})+i}{2\sqrt{2}}\right)^{\ell}
        \begin{pmatrix}
            z & -\overline{w}\\
            w & \overline{z}
        \end{pmatrix}
    \end{equation}
    with $z,w \in \bbZ[\sqrt{2}, i]$, $0\leq \ell \leq 7$, and an integer $k$.
\end{proposition}

\begin{proof}
    If a unitary $U$ is of the form Eq.~\eqref{eq: canonical form of Clifford+T}, it is a Clifford$+T$ matrix by Proposition \ref{prop: canonical form of Clifford+T 1}.
    We prove the contrary.
    Assume that $U$ is a Clifford$+T$ matrix.
    Then, again by Proposition \ref{prop: canonical form of Clifford+T 1}, $U$ is of the form
    \begin{equation}
        U = \begin{pmatrix}
            z' & -\overline{w'} e^{i\phi}\\
            w' & \overline{z'} e^{i\phi}
        \end{pmatrix}
    \end{equation}
    with $z,w \in \bbZ[\frac{1}{\sqrt{2}}, i]$ and $\phi\in \bbR$.
    Note that since $e^{i \phi} = \det U$, it is in the ring $\bbZ[i,1/\sqrt{2}]$.
    In fact, such a number is only a power of $\zeta_8$.
    Thus, we have
    \begin{equation}
        U = \begin{pmatrix}
            z' & -\overline{w'}\\
            w' & \overline{z'}
        \end{pmatrix}
        T^\ell
    \end{equation}
    for some $0\leq \ell \leq 7$.
    Let $k$ be the smallest denominator exponent of $UT^{-\ell}$.
    Since we have
    \begin{align}
        T &=
        \begin{pmatrix}
            1 & 0\\
            0 & \zeta_8
        \end{pmatrix}
        = \zeta_{16}
        \begin{pmatrix}
            \zeta_{16}^{-1} & 0\\
            0 & \zeta_{16}
        \end{pmatrix}\\
        &= \frac{(1 + \sqrt{2}) + i}{2\sqrt{2}}
        \begin{pmatrix}
            1-(1-\sqrt{2})i & 0\\
            0 & 1 + (1-\sqrt{2})i
        \end{pmatrix},
    \end{align}
    the unitary matrix $U$ is of the form Eq.~\eqref{eq: canonical form of Clifford+T} with $z = \sqrt{2}^k(1-(1-\sqrt{2})i)^{\ell} z'$ and $w = \sqrt{2}^k (1 + (1-\sqrt{2})i)^{\ell}w'$.
\end{proof}

\subsubsection{Clifford+V matrices}\label{subsec: Clifford+V}
    \begin{definition}
        For a $2\times 2$ matrix $A$, let $V_A = \frac{1}{\sqrt{5}} (I + 2iA)$.
        Let
        \begin{equation}
            X = \begin{pmatrix} 0 & 1 \\ 1 & 0\end{pmatrix},\quad
            Y = \begin{pmatrix} 0 & -i \\ i & 0\end{pmatrix},\quad
            Z = \begin{pmatrix} 1 & 0 \\ 0 & -1\end{pmatrix}.
        \end{equation}
        Clifford$+V$ operator is a unitary matrix given by a finite product of $\zeta_8I, S, H, V_X, V_Y, V_Z, V_X^{\dag}, V_Y^{\dag}, V_Z^{\dag}$.
    \end{definition}

    \begin{definition}
        For $U\in U(2)$, its $V$-count $\C(U,V,0)$ is defined as
        \begin{equation}
            \C(U,V,0):=\min\left\{N\in\mathbb{N}\ \middle|\
            \begin{array}{l}
                d(U,g_1g_2\cdots g_m)=0 \text{ for some Clifford$+V$ gates }g_1,\ldots, g_m,\\
                \text{with }m\geq 1,\ \text{ and } \#\{i\ |\ g_i = V\} = N
            \end{array}
            \right\}.
        \end{equation}
        If the set is empty, we define $\C(U,V,0)=\infty$.
    \end{definition}

    \begin{proposition}[{\cite[Proposition 7]{ross2014optimal}}]\label{prop: canonical form of Clifford+V 1}
        A unitary matrix $U \in U(2)$ is Clifford$+V$ matrix if and only if $U$ is of the form
        \begin{equation}\label{eq: canonical form of Clifford+V 1}
            U = 
            \frac{1}{\sqrt{5}^k \sqrt{2}^{\ell}}
            \begin{pmatrix}
                a & b\\
                c & d
            \end{pmatrix}
        \end{equation}
        with $a, b, c, d \in \bbZ[i]$, $0\leq \ell\leq 2$, and an integer $k$ such that $\det U$ is a power of $i$.
        Moreover, if $U$ is a Clifford$+V$ matrix, the minimum value of $k$ for all representations of $U$ in the form of Eq.~\eqref{eq: canonical form of Clifford+V 1} coincides with $\C(U,V,0)$.
    \end{proposition}

    \begin{proposition}\label{prop: canonical form of Clifford+V 2}
        A unitary matrix $U \in U(2)$ is Clifford$+V$ matrix if and only if $U$ is of the form
        \begin{equation}\label{eq: canonical form of Clifford+V 2}
            \frac{1}{\sqrt{5}^k \sqrt{2}^{\ell} (1-i)^m}
            \begin{pmatrix}
                z & -\overline{w}\\
                w & \overline{z}
            \end{pmatrix}
        \end{equation}
        with $z,w \in \bbZ[i]$, $0\leq \ell\leq 2$, $0\leq m\leq 3$ and an integer $k$ such that $\det U$ is a power of $i$.
        Moreover, if $U$ is a Clifford$+V$ matrix, the minimum value of $k$ for all representations of $U$ in the form of Eq.~\eqref{eq: canonical form of Clifford+V 1} coincides with $V(U)$.
    \end{proposition}

    \begin{proof}
        At first, we remark that the equality
        \begin{equation}
            S = \begin{pmatrix}
                1 & 0\\
                0 & i
            \end{pmatrix}
            = \frac{1}{1-i}
            \begin{pmatrix}
                1-i & 0\\
                0 & 1+i
            \end{pmatrix}
        \end{equation}
        holds.
        If a unitary matrix $U$ is of the form Eq.~\eqref{eq: canonical form of Clifford+V 2}, the matrix $US^{-m}$ is of the form Eq.~\eqref{eq: canonical form of Clifford+V 1}. Thus, the matrix $US^{-m}$ is Clifford$+V$ by Proposition \ref{prop: canonical form of Clifford+V 1}, so is $U$.

        We prove the contrary. Assume that $U$ is a Clifford$+V$.
        Then, the matrix $U$ is of the form Eq.~\eqref{eq: canonical form of Clifford+V 1} by Proposition \ref{prop: canonical form of Clifford+V 1}.
        Since $U$ is a unitary matrix, it is written as
        \[
            \frac{1}{\sqrt{5}^k \sqrt{2}^\ell}
            \begin{pmatrix}
                z' & -\overline{w'} e^{i\phi}\\
                w' & \overline{z'} e^{i\phi}
            \end{pmatrix}
        \]
        with $z', w' \in \bbZ[i]$ and $\phi \in \bbR$.
        The equality $e^{i\phi} = \det U = ad - bc$ implies that $e^{i\phi}$ is in the ring $\bbZ[i]$.
        Every element of $\bbZ[i]$ with the absolute value $1$ is some power of $i$.
        Thus, $U$ is of the form
        \begin{align}
            U &= \frac{1}{\sqrt{5}^k \sqrt{2}^\ell}
            \begin{pmatrix}
                z' & -\overline{w'} i^n\\
                w' & \overline{z'} i^n
            \end{pmatrix}
            =
            \frac{1}{\sqrt{5}^k \sqrt{2}^\ell}
            \begin{pmatrix}
                z' & -\overline{w'}\\
                w' & \overline{z'}
            \end{pmatrix}
            S^n\\
            &= \frac{1}{\sqrt{5}^k \sqrt{2}^\ell (1-i)^n}
            \begin{pmatrix}
                z' & -\overline{w'}\\
                w' & \overline{z'}
            \end{pmatrix}
            \begin{pmatrix}
                1-i & 0\\
                0   & 1+i
            \end{pmatrix}^n.
        \end{align}
        Letting $z = z'(1-i)^n$ and $w = w'(1+i)^n$, we get the assertion.
        The last statement is a consequence of Proposition \ref{prop: canonical form of Clifford+V 1}.
    \end{proof}

\subsubsection{Arithmetic quantum matrices and generalized counts}\label{subsec: arithmetic quantum}
\begin{definition}
    Let $K$ be a totally real number field, i.e., a number field such that all its embeddings into $\bbC$ have the image in $\bbR$.
    Let $\calO_K$ be the ring of integers of $K$.
    Let $M_K$ (resp. $M_K^{\infty}$) be the set of standard absolute values (resp. standard non-Archimedean absolute values) defined in Section \ref{subsec: absolute values}.
    Let $S\subset M_K$ be a finite set containing $M_K^{\infty}$, and $\calX$ be a finite set of algebraic numbers.
    We call $(K, S,\calX)$ an arithmetic datum.
    We say that a unitary matrix $V \in U(2)$ is {\it arithmetic quantum gate} for the arithmetic datum $(K, S, \calX)$ if $V$ is of the form
    \begin{equation}\label{eq: AQ}
        V = \frac{1}{u_1 u_2}
        \begin{pmatrix}
            \alpha + i\beta & -\gamma + i\delta\\
            \gamma + i\delta & \alpha - i\beta
        \end{pmatrix}
    \end{equation}
    with $\alpha,\beta, \gamma, \delta, u_1 \in\calO_K$ and $u_2 \in \calX$ such that
    \[
    \Supp(u_1) \coloneqq \left\{ v\in M_K\ \middle|\ |u_1|_v \neq 1\right\}
    \]
    is contained in $S$.
    We denote $\AQ(K, S,\calX)$ for the set of the arithmetic quantum matrices for $(K, S,\calX)$.
\end{definition}

\begin{definition}
    Let $(K, S,\calX)$ be an arithmetic datum.
    For $V \in \AQ(K,S,\calX)$, the {\it lowest denominator height} $\LDH_{K,S,\calX}(V)$ is defined by
    \begin{equation}
        \LDH_{K,S,\calX}(V) \coloneqq \min \left\{ H_K(u_1)\ \middle|\ 
        \begin{array}{l}
            \alpha, \beta, \delta,\gamma, u_1 \in \calO_K, u_2 \in \calX \\  \text{ satisfy }Eq.~\eqref{eq: AQ} \text{ and }\Supp(u_1)\subset S
        \end{array}
        \right\},
    \end{equation}
    where $H_K$ is the naive relative height defined in Definition \ref{dfn: height function}
    For $U\in U(2)$, $\scrC \subset \AQ(K,S,\calX)$, and $\varepsilon >0$,
    the approximated lowest denominator height $\LDH_{\scrC}(U,\varepsilon)$ is defined by
    \begin{equation}
        \LDH_{\scrC}(U, \varepsilon) \coloneqq \min \left\{ \LDH_{K,S,\calX}(V)\ \middle|\ V\in \scrC, d(U,V) < \varepsilon\right\}.
    \end{equation}
    If there is no such $V\in \scrC$, we define $\LDH_{\scrC}(U,\varepsilon) = +\infty$ for convenience.
    The upper (resp. lower) logarithmic order of the lowest denominator height $\overline{\ldh}_{\scrC}(U)$ (resp. $\underline{\ldh}_{\scrC}(U)$) is defined by
    \begin{align}
        \ldhupper_{\scrC}(U) &\coloneqq \inf\left\{t\in\bbR\  \middle| \ \exists \varepsilon_0 > 0, \forall \varepsilon \in (0,\varepsilon_0),\ \LDH_{\scrC}(U,\varepsilon) \leq \left(\frac{1}{\varepsilon}\right)^{t}\right\},\\
        \ldhlower_{\scrC}(U) &\coloneqq \sup\left\{t\in\bbR\  \middle| \ \exists \varepsilon_0 > 0, \forall \varepsilon \in (0,\varepsilon_0),\ \LDH_{\scrC}(U,\varepsilon) \geq \left(\frac{1}{\varepsilon}\right)^{t}\right\}.
    \end{align}
\end{definition}

\subsubsection{How to use Theorem \ref{thm: ldhlower geq 2}} \label{subsec: how to use}\hphantom{=}

In this section, we explain how Theorem \ref{thm: ldhlower geq 2} implies Theorem \ref{thm: G-lower geq 4}.

\begin{proof}[Theorem \ref{thm: ldhlower geq 2} implies Theorem \ref{thm: G-lower geq 4} (i)]
    By Proposition \ref{prop: canonical form of Clifford+T 2}, the set of Clifford$+T$ matrices is equal to $\AQ(\bbQ(\sqrt{2}), S, \calX)$ with
    \begin{align}
        S &= M_{\bbQ(\sqrt{2})}^{\infty}\cup\{\sqrt{2}\bbZ[\sqrt{2}]\}, \text{ and}\\
        \calX &= \left\{ \left(\frac{(1+\sqrt{2})+i}{2\sqrt{2}}\right)^{-\ell}\ \middle|\ 0\leq \ell \leq 7\right\}.
    \end{align}
    An element $u_1\in \bbZ[\sqrt{2}]$ satisfies $\Supp(u_1) \subset S$ if and only if $u_1$ is of the form
    \begin{equation}
        u_1 = \pm \sqrt{2}^k(1+\sqrt{2})^i
    \end{equation}
    for some integer $k\geq 0$ and $i \in \bbZ$.
    Since we have the equality
    \begin{equation}
        H_{\bbQ(\sqrt{2})}(\pm \sqrt{2}^k(1+\sqrt{2})^i) =
        \begin{cases}
            \sqrt{2}^k(1+\sqrt{2})^i (> 2^k) &\text{if } |\sqrt{2}^k(1-\sqrt{2})^i| < 1,\\
            2^k &\text{otherwise,}
        \end{cases}
    \end{equation}
    the quantity $\LDH_{\bbQ(\sqrt{2}),S,\calX}(V)$ is equal to $2^{\lde(V,\sqrt{2})}$.
    Moreover, the $T$-count of $V$ is at least $2\lde(V,\sqrt{2})-3$~\cite{giles2019remarksmatsumotoamanosnormal}.
    Hence, we conclude that Theorem \ref{thm: ldhlower geq 2} implies Theorem \ref{thm: G-lower geq 4} (i).
\end{proof}

\begin{proof}[Theorem \ref{thm: ldhlower geq 2} implies Theorem \ref{thm: G-lower geq 4} (ii)]
    By Proposition \ref{prop: canonical form of Clifford+V 2}, the set of Clifford$+V$ matrices is equal to $\AQ(\bbQ, S, \calX)$ with
    \begin{align}
        S &= M_{\bbQ}^{\infty}\cup \{5\bbZ \}, \text{ and}\\
        \calX &= \left\{ \sqrt{5}^{k_5}\sqrt{2}^{k_2} (1-i)^{k_0} \middle|\ 0\leq k_5 \leq 1, 0\leq k_2 \leq 2, 0\leq k_0 \leq 3\right\}.
    \end{align}
    An element $u_1\in \bbZ$ satisfies $\Supp(u_1) \subset S$ if and only if $u_1$ is of the form
    \begin{equation}
        u_1 = \pm 5^k
    \end{equation}
    for some integer $k\geq 0$.
    Since we have the equality
    \begin{align}
        H_{\bbQ}(\pm 5^k) = 5^k,
    \end{align}
    the quantity $\LDH_{\bbQ(\sqrt{2}),S,\calX}(V)$ is equal to $5^{\lfloor\lde(V,\sqrt{5})/2\rfloor}$, where we let
    \[
        \lde(V,\sqrt{5}) \coloneqq \min\left\{ k\in \Z\ \middle|\ V = \frac{1}{\sqrt{5}^k \sqrt{2}^{\ell}}
        \begin{pmatrix}
            a & b\\
            c & d
        \end{pmatrix} \text{ with } k,\ell \in \Z, a,b,c,d \in \Z[i]\right\}.
    \]
    Moreover, the $V$-count of $V$ is equal to $\lde(V,\sqrt{5})$.
    Hence, we conclude that Theorem \ref{thm: ldhlower geq 2} implies Theorem \ref{thm: G-lower geq 4} (ii).
\end{proof}

\subsection{Proof of Theorem \ref{thm: ldhlower geq 2}}\label{sec: Proof of main theorem}
In this final section, we prove Theorem \ref{thm: ldhlower geq 2}.
\begin{proof}[Proof of Theorem \ref{thm: ldhlower geq 2}]
    Let $(a,b,c,d)=\frac{1}{L}(\alpha,\beta,\gamma,\delta)$ with $\alpha,\beta,\gamma,\delta\in \calO_K$ and $L = \sqrt{\alpha^2+\beta^2+\gamma^2+\delta^2}$.

    For $\varepsilon > 0$, suppose that $V=\frac{1}{u_1 u_2}\begin{pmatrix}
        \alpha'+i\beta'&-\gamma'+i\delta'\\
        \gamma'+i\delta'&\alpha'-i\beta'
    \end{pmatrix} \in \scrC$
    with $u_1,\alpha',\beta',\gamma',\delta'\in\calO_K$, $u_2 \in \overline{\bbQ}$, $(u_1 u_2)^2 = \alpha'^2+\beta'^2+\gamma'^2+\delta'^2$, and $\Supp(u_1) \subset S$ satisfies
    \begin{equation} \label{eq:approximation of matrix}
        0 < d(U,V)\leq\varepsilon, \quad
        H_K(u_1) = \LDH_{\scrC}(U,\varepsilon).
    \end{equation}
    Inequality Eq.~\eqref{eq:approximation of matrix} implies that
    \begin{equation}\label{eq: bound of main absolute value}
        |u_2L|_{\sigma_1}(1-\varepsilon^2)<|u_2L|_{\sigma_1}\sqrt{1-\varepsilon^2}\leq\frac{|\lambda|_{\sigma_1}}{|u_1|_{\sigma_1}}< |u_2L|_{\sigma_1},
    \end{equation}
    where $\lambda=\alpha\alpha'+\beta\beta'+\gamma\gamma'+\delta\delta'\in \calO_K$.
    For $i=1,2,\ldots, [K:\bbQ]$, let
    \begin{align}
        \sigma_i(L) &\coloneqq \sqrt{\sigma_i(L^2)},\\
        \sigma_i(U) &\coloneqq \frac{1}{\sigma_i(L)}
        \begin{pmatrix}
            \sigma_i(\alpha) + i\sigma_i(\beta) & -\sigma_i(\gamma) + i\sigma_i(\delta)\\
            \sigma_i(\gamma) + i\sigma_i(\delta) & \sigma_i(\alpha) - i\sigma_i(\beta)
        \end{pmatrix},\\
        \sigma_i(V) &\coloneqq \frac{1}{\sigma_i(u_1 u_2)}
        \begin{pmatrix}
            \sigma_i(\alpha') + i\sigma_i(\beta') & -\sigma_i(\gamma') + i\sigma_i(\delta')\\
            \sigma_i(\gamma') + i\sigma_i(\delta') & \sigma_i(\alpha') - i\sigma_i(\beta')
        \end{pmatrix}.\\
    \end{align}
    Since $\sigma_i(U)$ and $\sigma_i(V)$ are single-qubit unitary operators, $|\tr{\sigma_i(U)^\dag \sigma_i(V)}|\leq2$ holds, and it implies
     \begin{equation}\label{eq: bound of other absolute values}
        \frac{|\lambda|_{\sigma_i}}{|u_1|_{\sigma_i}}\leq |u_2 L|_{\sigma_i}.
    \end{equation}

    Assume the inequality $\ldhlower_{K,S,\calX}(U)<2$ and write $\ldhlower_{K,S,\calX}(U) = 2 - 2\mu$ with $\mu>0$.
    Then, for all $\varepsilon_0>0$, there exists $\varepsilon\in(0,\varepsilon_0)$ such that the inequality \begin{equation}\label{eq: assumption of LDH}
        \LDH_{\scrC}(U,\varepsilon)<\left(\frac{1}{\varepsilon}\right)^{2-\mu}
    \end{equation}
    holds.
    Take an infinite sequence $\varepsilon_1 > \varepsilon_2 > \cdots > 0$ such that each $\varepsilon_j$ satisfies Eq.~\eqref{eq: assumption of LDH} with $\varepsilon = \varepsilon_j$.
    The inequality Eq.~\eqref{eq: assumption of LDH} is equivalent to the existence of $V_{\varepsilon}=\frac{1}{u_1 u_2}\begin{pmatrix}
        \alpha'+i\beta'&-\gamma'+i\delta'\\
        \gamma'+i\delta'&\alpha'-i\beta'
    \end{pmatrix} \in \AQ(K,S,\calX)$
    with $u_1,\alpha', \beta', \gamma', \delta' \in \calO_K$, $u_2 \in \calX$, and $\Supp(u_1) \subset S$ satisfying the inequalities
    \begin{align}
        d(U,V_{\varepsilon}) &< \varepsilon, \text{ and}\\
        H_K(u_1) &= \LDH_{K,S,\calX}(V_{\varepsilon}) < \left( \frac{1}{\varepsilon}\right)^{2-\mu}.
    \end{align}
    By combining this and Eq.~\eqref{eq: bound of main absolute value}, we obtain
    \begin{equation}\label{eq: height of u1}
        0< |u_2L|_{\sigma_1} - \left| \frac{\lambda}{u_1}\right|_{\sigma_1} <\frac{|u_2L|_{\sigma_1}}{H_K(u_1)^{2/(2-\mu)}}.
    \end{equation}
    We identify $K$ with its image $\sigma_1(K)$.
    Since $K$ is a totally real field, the value $|\lambda/u_1|_{\sigma_1}$, which is either $\lambda_1/u_1$ or $-\lambda_1/u_1$, is itself an element of $K$.
    Moreover, since we have $|u_2L|_{\sigma_1}^2 = \sigma_1(u_2L) \cdot \overline{\sigma_1(u_2L)}$, the value $|u_2L|_{\sigma_1}$ is an algebraic number.
    Put $c = 2/(2-\mu)>1$,  $A = |u_2 L|_{\sigma_1}$, and $C_i = |u_2L|_{\sigma_i}$ $(1\leq i \leq [K:\bbQ])$.
    Then, by Proposition \ref{prop: approximation}, we see that for each $u_2 \in \calX$, the number of $|\lambda/u_1|_{\sigma_1}$ satisfying Eq.~\eqref{eq: bound of other absolute values} and Eq.~\eqref{eq: height of u1} is finite.

    Again, by using Eq.~\eqref{eq: height of u1}, the value $H_K(u_1)$ is bounded above by the maximum of the value $|u_2L|_{\sigma_1}/(|u_2L|_{\sigma_1} - |\lambda/u_1|_{\sigma_1})$ along all candidates of $\lambda/u_1$.
    Hence, the number of candidates of $u_1$ is at most finite by Theorem \ref{thm: Northcott property}.

    Let $\xi$ be $\alpha', \beta', \gamma',$ or $\delta'$.
    Then, for each $1\leq i \leq [K:\bbQ]$, we have the inequalities
    \begin{align}
        \|\xi\|_{\sigma_i}^2 \leq \|\alpha'^2 + \beta'^2 + \gamma'^2 + \delta'^2\|_{\sigma_i} = \|u_1 u_2\|_{\sigma_i}^2.
    \end{align}
    Taking $\prod_{i=1}^{[K:\bbQ]}\max\{1, |\cdot|_{\sigma_i}\}$, we obtain the upper bound of $H_K(\xi)$.
    Consequently, the number of candidates of $\xi$ is at most finite by Theorem \ref{thm: Northcott property}.   
    Thus, the set $\{V_{\varepsilon_j}\ |\ j = 1,2,\ldots \}$ is finite.
    This is equivalent to that $d(U,V_{\varepsilon_j}) = 0$ for sufficiently large $j$.
    But since $\calU$ is not realized by elements of $\scrC$ by assumption, this is a contradiction.
\end{proof}

\section*{Acknowledgements}
HM is supported by 
MEXT Q-LEAP Grant No. JPMXS0120319794 and JST SPRING Grant No. JPMJSP2138.
SA is partially supported by JST PRESTO Grant no.JPMJPR2111, JST Moonshot R\&D MILLENNIA Program (Grant no.JPMJMS2061), JPMXS0120319794, and CREST (Japan Science and Technology Agency) Grant no.JPMJCR2113.
\begin{appendices}

\end{appendices}

\bibliographystyle{plain}


\end{document}